\documentclass[acmtog,nonacm,screen]{acmart}

\usepackage[bbgreekl]{mathbbol}

\usepackage{enumitem}
\usepackage{gensymb}
\usepackage{multirow}
\usepackage{prettyref}
\usepackage[labelfont=bf,format=plain,singlelinecheck=false]{caption}
\usepackage{subcaption}
\usepackage{bm,microtype}
\usepackage{overpic}
\definecolor{myblue}{RGB}{31, 119, 180}
\definecolor{myorange}{RGB}{255, 127, 14}
\definecolor{mygreen}{RGB}{44, 160, 44}
\definecolor{myred}{RGB}{214, 39, 40}

\newcommand{\todo}[1]{\textcolor[RGB]{255, 0, 0}{[\textbf{TODO:} {\em #1}]}}

\usepackage{soul}
\AtBeginDocument{%
  }

\setcopyright{acmlicensed}
\acmJournal{TOG}

\usepackage{tikz}
\usetikzlibrary{angles, calc}
\tikzset{
  dot/.style={
    circle, fill=black, inner sep=1pt, outer sep=0pt
  },
  dot label/.style={
    circle, inner sep=0pt, outer sep=1pt
  },
  pics/right angle/.append style={
    /tikz/draw, /tikz/angle radius=5pt
  }
}
\usepackage{xcolor}

\acmSubmissionID{373}

\usepackage[noend]{algpseudocode}

\def\BState{\State\hskip-\ALG@thistlm}
\makeatother

\newcommand{\algorithmicforeach}{\textbf{foreach}}

\algdef{SE}[FOREACH]{ForEach}{EndForEach}[1]{\algorithmicforeach\ #1\ \algorithmicdo}{\algorithmicend\ \algorithmicforeach}%
\algtext*{EndForEach}

\usepackage{prettyref}
\newrefformat{fig}{Figure~\ref{#1}}
\newrefformat{par}{Section~\ref{#1}}
\newrefformat{appen}{Appendix~\ref{#1}}
\newrefformat{sec}{Section~\ref{#1}}
\newrefformat{sub}{Section~\ref{#1}}
\newrefformat{table}{Table~\ref{#1}}
\newrefformat{ass}{Assumption~\ref{#1}}
\newrefformat{alg}{Algorithm~\ref{#1}}
\newrefformat{def}{Definition~\ref{#1}}
\newrefformat{thm}{Theorem~\ref{#1}}
\newrefformat{cor}{Corollary~\ref{#1}}
\newrefformat{lem}{Lemma~\ref{#1}}
\newrefformat{step}{Step~\ref{#1}}
\newrefformat{ln}{Line~\ref{#1}}
\newrefformat{rem}{Remark~\ref{#1}}
\newrefformat{eq}{Equation~\ref{#1}}
\newrefformat{pb}{Problem~\ref{#1}}
\newrefformat{it}{Item~\ref{#1}}
\newrefformat{te}{Term~\ref{#1}}
\def\Eqref Eq:#1:{\eqref{eq:#1}}
\newrefformat{Eq}{Equation~\Eqref#1:}

\usepackage{rotating}

\newcommand{\bbx}{\mathbb{x}}

\newcommand{\bbv}{\mathbb{v}}
\newcommand{\bbu}{\mathbb{u}}
\newcommand{\bbg}{\mathbb{g}}
\newcommand{\bbf}{\mathbb{f}}
\newcommand{\bbF}{\mathbb{F}}
\newcommand{\bbP}{\mathbb{P}}
\newcommand{\bbp}{\mathbb{p}}
\newcommand{\bbq}{\mathbb{q}}
\newcommand{\bbC}{\mathbb{C}}
\newcommand{\bbw}{\mathbb{w}}
\newcommand{\bbm}{\mathbb{m}}
\newcommand{\bbt}{\mathbb{t}}

\newcommand{\uu}{\mathbf{u}}

\newcommand{\vvv}{\mathbf{v}}

\newcommand{\cc}{\mathbf{c}}

\renewcommand{\SS}{\mathbf{S}}

\newcommand{\FF}{\mathbf{F}}

\newcommand{\sig}{\boldsymbol\sigma}

\newcommand{\px}[2]{\frac{\partial{}#1}{\partial{}#2}}

\newcommand\restr[2]{{\left.\kern-\nulldelimiterspace{}#1\right|_{#2}}}
\newcommand{\tr}{\mathrm{tr}}

\newcommand{\refine}{{\fontfamily{lmss}\selectfont \textcolor[RGB]{160, 32, 240}{refine}}}
\newcommand{\delete}{{\fontfamily{lmss}\selectfont \textcolor[RGB]{160, 32, 240}{delete}}}
\newcommand{\border}{{\fontfamily{lmss}\selectfont \textcolor[RGB]{0, 191, 0}{border}}}
\newcommand{\leaf}{{\fontfamily{lmss}\selectfont \textcolor[RGB]{0, 191, 0}{leaf}}}
\newcommand{\coarsen}{{\fontfamily{lmss}\selectfont \textcolor[RGB]{160, 32, 240}{coarsen}}}

\usepackage{float}
\usepackage{algorithm} 
\usepackage{algpseudocode} 
\usepackage{svg}
\usepackage{xcolor}
\usepackage{wrapfig}
\usepackage[export]{adjustbox}
\usepackage{colortbl}
\usepackage{tikz}
\usetikzlibrary{angles, calc}
\tikzset{
  dot/.style={
    circle, fill=black, inner sep=1pt, outer sep=0pt
  },
  dot label/.style={
    circle, inner sep=0pt, outer sep=1pt
  },
  pics/right angle/.append style={
    /tikz/draw, /tikz/angle radius=5pt
  }
}

\definecolor{smGrey}{rgb}{0.8, 0.8, 0.8}  
\definecolor{smOrange}{rgb}{1.0, 0.64, 0.0}  
\definecolor{smBlue}{rgb}{0.0, 0.64, 1.0}

\begin{document}

\title{Adaptive GPU Kinetic Solver for Fluid–Granular Flows}

\author{Xingqiao Li}
\authornote{This work was done when Xingqiao Li was an intern at LIGHTSPEED.}
\email{lixingqiao@pku.edu.cn}
\affiliation{
\institution{Peking University}
\country{China}
}

\author{Kui Wu}
\email{walker.kui.wu@gmail.com}
\affiliation{
\institution{LIGHTSPEED}
\country{USA}
}

\author{Haozhe Su}
\email{dyhard0520@gmail.com}
\affiliation{
\institution{LIGHTSPEED}
\country{USA}
}

\author{Tianhong Gao}
\email{guesss2022@gmail.com}
\affiliation{
\institution{Peking University}
\country{China}
}

\author{Mengyu Chu}
\email{mchu@pku.edu.cn}
\affiliation{
\institution{Peking University}
\country{China}
}

\author{Chenfanfu Jiang}
\email{Chenfanfu.Jiang@gmail.com}
\affiliation{
\institution{UCLA}
\country{USA}
}

\author{Wei Li}
\email{1104720604wei@gmail.com}
\affiliation{
\institution{Shanghai Jiao Tong University}
\country{China}
}

\author{Baoquan Chen}
\email{baoquan@pku.edu.cn}
\affiliation{
\institution{Peking University}
\country{China}
}

\begin{abstract}
Simulating fluid–granular flows is crucial for understanding natural disasters, industrial processes, and visually realistic phenomena in computer graphics. These systems are challenging to simulate because of the strong nonlinear coupling between continuum fluids and discrete granular media, making it difficult to achieve both physical fidelity and computational efficiency at large scales. In this work, we present a unified framework for large-scale fluid–granular simulation that couples the Lattice Boltzmann Method (LBM) for fluids with the Material Point Method (MPM) for granular materials such as sand and snow. We introduce an adaptive block-based multi-level HOME-LBM solver based on solid geometric structures, enabling efficient memory usage and computational performance across multiple lattice resolutions. Consistent rescaling laws for moments allow accurate transfer of macroscopic quantities across refinement interfaces, while a GPU-based algorithm dynamically maintains the multi-level blocks in response to particle motion. By enforcing that all MPM particles reside within the finest fluid nodes, we achieve accurate two-way coupling between fluid and granular phases. Our framework supports a wide range of large-scale phenomena, including snow avalanches, sandstorms, and sand migration, demonstrating high physical fidelity and computational efficiency.


\end{abstract}

\ccsdesc[500]{Computing methodologies~Physical simulation}

%
%

\keywords{LBM, MPM, GPU, Adaptive grid}

\begin{teaserfigure}
\includegraphics[trim = 100 50 100 0, clip, width=0.495\textwidth]{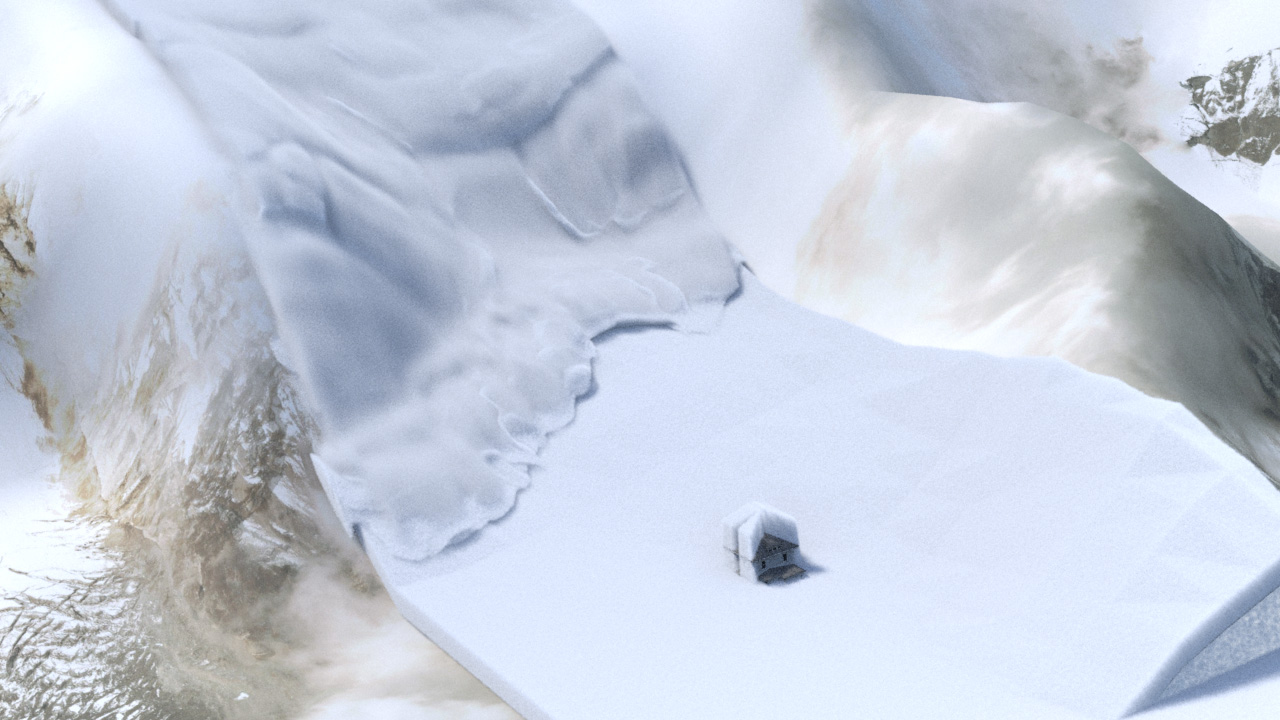}
\includegraphics[trim = 100 50 100 0, clip, width=0.495\textwidth]{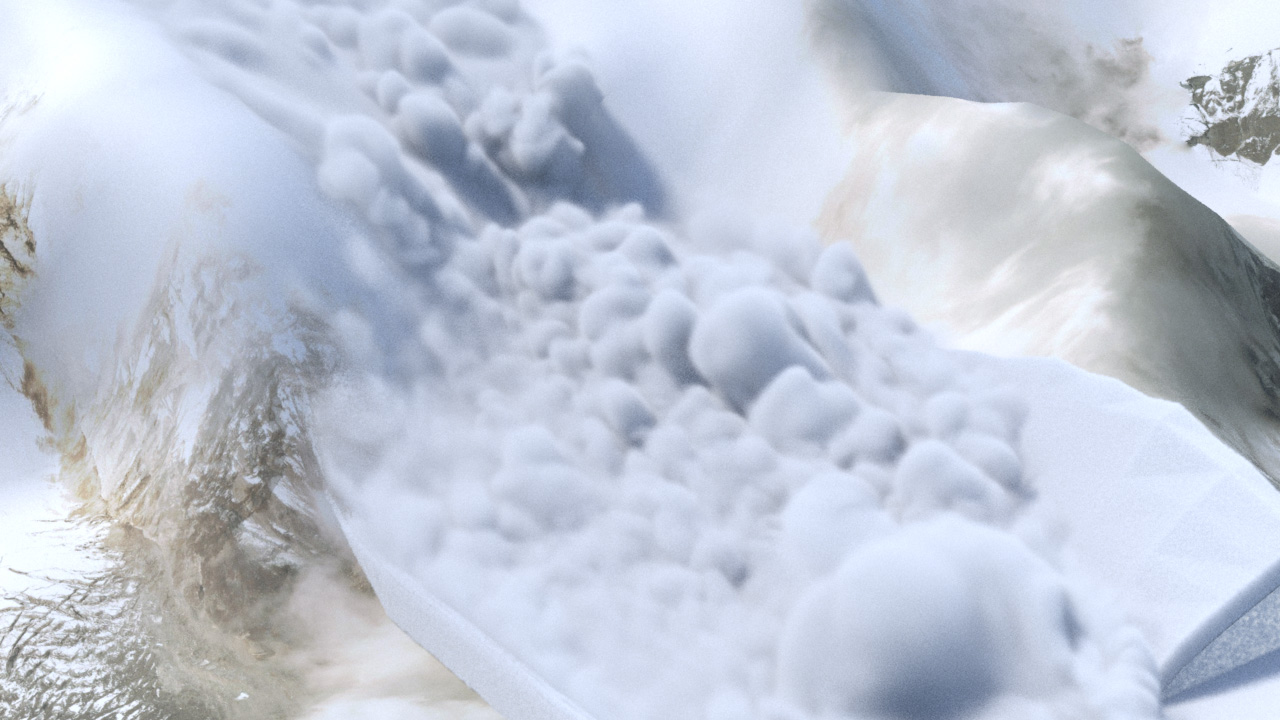}
  \caption{A large-scale snow avalanche on a mountain, producing the effect of turbulent powder snow clouds with the interaction of MPM snow and LBM air. Simulated on GPU in an effective LBM grid resolution of $1536\times 768\times 384$ and 55.8M MPM particles.}
  \Description{}
  \label{fig:teaser}
\end{teaserfigure}

\maketitle




\section{Introduction}

Simulating fluid–granular flows is critically important, as these phenomena are ubiquitous, high-impact, and difficult to study experimentally, yet directly affect safety, infrastructure, industry, and visual realism. Many natural disasters, including snow avalanches, sandstorms, mudslides, and tsunamis, involve strong fluid–granular interactions, which are rare, destructive, and unsafe to reproduce in controlled settings, making simulation an essential tool for analysis and prediction. Fluid–granular systems are inherently challenging to simulate because they couple two fundamentally different physical systems, continuum fluids and discrete granular media, through strong, nonlinear interactions. Achieving both physical fidelity and computational efficiency at large scales remains difficult.

In computer graphics, fluid simulation has advanced through Eulerian~\cite{Stam1999, Qu2019, Sun2025Leapfrog} and hybrid Eulerian–Lagrangian methods~\cite{Jiang2015, Zhang2016, Fu2017, Nabizadeh2024}, while particle-based approaches, including the Material Point Method (MPM)~\cite{Klar2016Sand, Daviet2016mpmsand, Stomakhin2013, Yu2024}, smoothed particle hydrodynamics (SPH)~\cite{benes06, Alduan2011SPHGranular, Takahashi2021sphrigid, Gissler2020}, and the discrete element method (DEM)~\cite{Bell2005Granular, Yue2018HybridGrains}, are widely used for granular materials due to their robustness and ability to capture large deformations and plasticity. Coupling fluids and granular media remains a challenge. Various strategies have been proposed~\cite{Lenaerts2009sphwatersand, Yan2016MultiphaseSPH, Yang2017UnifiedParticle, Ren2021UnifiedParticle, Tampubolon2017MultiSpecies, Gao2018Mixture, su2023, Rungjiratananon2008SandWater, Wang2021sphdem, Tang2025, Nascimento2025}. Despite these advances, existing methods remain limited in both physical fidelity and computational efficiency for large-scale simulations.

In this work, we derive a locally averaged fluid formulation for particle-laden flows within a single-phase Lattice Boltzmann framework. We model the fluid using HOME-LBM~\cite{li2023high} and granular materials, including sand and snow, using MPM. To enable large-scale coupled HOME-LBM–MPM simulations, we propose an adaptive block-based multi-level HOME-LBM solver that achieves both memory and computational efficiency, along with an efficient parallel grid construction and update scheme that enables rapid adaptation to complex geometric structures.
We derive consistent rescaling laws for moment variables across lattice levels and integrate them into an advanced multi-level  HOME-LBM scheme, enabling efficient transfer of macroscopic quantities across refinement interfaces. 
To ensure accurate two-way coupling, we enforce that all MPM particles reside within LBM nodes at the finest level and introduce a GPU-based algorithm to dynamically maintain the multi-level fluid grid in response to particle motion. 
Our framework enables various large-scale phenomena, including snow avalanches, sandstorms, and sand migration.


\section{Related works}

In this section, we provide a brief overview of previous simulation methods for wind, fluids, snow, sand, and their mixtures. 

\paragraph{Fluid Dynamics}
Fast, high-fidelity fluid simulation has long been a core challenge in computer graphics. Stam’s stable fluids~\cite{Stam1999} introduced an unconditionally stable Eulerian formulation for the Navier–Stokes equations, inspiring numerous improvements in advection accuracy and vortical detail, including BFECC~\cite{Kim2005}, modified MacCormack~\cite{Selle2008}, bidirectional and neural flow maps~\cite{Qu2019, Deng2023}, and Lagrangian vorticity treatments using particles, filaments, and sheets~\cite{Selle2005, Weismann2010, Pfaff2012, Zhang2015}. Hybrid Eulerian–Lagrangian methods, such as FLIP-based extensions~\cite{Jiang2015, Zhang2016, Fu2017, Nabizadeh2024}, further enhance accuracy by combining grids and particles. More recently, mesoscopic approaches like the lattice Boltzmann method (LBM)~\cite{Li2003, Thurey2009} have been applied to both single-phase~\cite{Wei2020, Lyu2021, li2023high, Liu2023, Liu2025} and multiphase flows~\cite{WeiTVCG2021, Wei2022, Wei2023, dinghybrid,xiao2025simulating, Wang2025lbmfoam}.


\paragraph{Granular Simulation}
Particle-based approaches model millions of interacting grains, with early methods representing inter-grain dynamics explicitly using pairwise interactions~\cite{luciani1995multi}. These interactions have been handled using the discrete element method (DEM)~\cite{Bell2005Granular}, smoothed particle hydrodynamics (SPH) for sand and snow~\cite{benes06, Alduan2011SPHGranular, Takahashi2021sphrigid, Gissler2020}, hybrid SPH–DEM formulations~\cite{Bell2005Granular}, and projective peridynamics~\cite{He2018Peridynamicssand}. Hybrid Lagrangian–Eulerian schemes, such as particle-in-cell (PIC) methods~\cite{Zhu2005fluidsand, Narain2010sand}, improve stability and accuracy, while the Material Point Method (MPM) further augments particles with attributes like the deformation gradient, enabling realistic large-scale simulations of sand~\cite{Klar2016Sand, Daviet2016mpmsand} and snow~\cite{Stomakhin2013, Yu2024}. MPM can also be combined with DEM to enhance efficiency in granular flow simulations~\cite{Yue2018HybridGrains}. Our work shares the core idea of using separate grids for fluids and grains with~\citet{Gao2018Mixture}, but differs in employing an adaptive LBM framework to simulate large-scale, highly turbulent fluid motion. In contrast to the layered abstraction of~\citet{Nascimento2025}, our method models dense granular layers with MPM and their coupling to the fluid more realistically.

\paragraph{Mixture}
One example of fluid–granular interaction is water–sand mixing, which can be modeled by treating water and sand as particles, phases, or materials and coupling them using SPH–DEM~\cite{Rungjiratananon2008SandWater, Wang2021sphdem} or unified SPH~\cite{Lenaerts2009sphwatersand, Yan2016MultiphaseSPH, Yang2017UnifiedParticle, Ren2021UnifiedParticle}. Beyond particle-based approaches, hybrid Eulerian–Lagrangian methods enable stronger two-way coupling. For instance, DEM–PIC formulations~\cite{Tang2025} and extensions of the Power PIC method~\cite{Qu2023} allow bidirectional interaction between water and sand, while two-grid MPM schemes~\cite{Tampubolon2017MultiSpecies} and dual-background grid methods~\cite{Gao2018Mixture} resolve momentum exchange between FLIP fluids and MPM sediments. To trade some physical fidelity for efficiency, shallow-water or layered abstractions have been proposed for two-phase sand–water flows~\cite{su2023} and powder-snow avalanches~\cite{Nascimento2025}, using height or layer fields to approximate inter-phase dynamics.

\paragraph{Efficient Implementation} 
A variety of CPU- and GPU-based strategies have been developed to accelerate large-scale simulation. These include fast Poisson solvers using multigrid~\cite{Chentanez2011, Sun2025Leapfrog}, fixed-point iterations~\cite{Chen2015}, or compact Poisson filters~\cite{Rabbani2022}; memory- and computation-efficient sparse structures~\cite{Wu2018gvdb, Hu2019Taichi, Gao2018gpumpm}, adaptive grids~\cite{Wang2025cirrus, Braun2025}, and quantization~\cite{Hu2021QuanTaichi, Liu2022autoquantization}; and parallelism via multi-GPU execution~\cite{Horvath2009, Wang2020}. In contrast, the lattice Boltzmann method (LBM) operates locally on lattice nodes at the mesoscopic scale, making it particularly well suited for GPU acceleration and large-scale flows~\cite{Rinaldi2012}. Its efficiency has been further enhanced through cache-friendly data layouts~\cite{Chen2022, lehmann2022esoteric}, moment-based representations~\cite{li2023high}, sparse domains~\cite{Wang2025}, and multiresolution grids~\cite{Lyu2023, li2018}.
Unfortunately, these LBM-based works are grid-based, multi-resolution, and do not support adaptive blocks or dynamic updates for moving objects, thereby limiting their applicability.


\section{Background}

This section provides a brief overview of the HOME-LBM~\cite{li2023high} method and the theory of macroscale particle-laden flow.

\subsection{Lattice Boltzmann Method (LBM)}

In the context of single-phase fluid dynamics, the evolution of the particle distribution function is governed by the lattice Boltzmann equation (LBE) in dimensionless units:
\begin{equation}
f_i(\bm{x}+\bm{c}_i, t+1) - f_i(\bm{x}, t) = \Omega_i + F_i \;,
\label{eq:lbe_normalized}
\end{equation}
where $f_i(\bm{x}, t)$ denotes the distribution function along the $i$-th lattice direction at position $\bm{x}$ and time $t$, $\bm{c}_i$ is the corresponding discrete lattice velocity, $\Omega_i$ represents the collision operator, and $F_i$ accounts for external forces projected into distribution space.
The evolution in~\autoref{eq:lbe_normalized} is solved using an operator-splitting scheme with two steps. First, the distribution functions are advected through a streaming step,
followed by a collision step,

The macroscopic quantities $\mathcal{M}={\rho,\bm{u},\bm{S}}$, density $\rho$, velocity $\bm{u}$, and the second-order moment tensor $\bm{S}$ related to the stress—are recovered from the distribution functions as
\begin{equation}
\rho=\sum_{i=0}^{q-1} f_i, \; \rho\bm{u} = \sum_{i=0}^{q-1}\bm{c}_i f_i + \frac{1}{2}\bm{F}, \;
\rho {S}_{\alpha \beta}=\sum_{i=0}^{q-1}(\bm{c}_{i \alpha}\bm{c}_{i \beta} -c_s^2\delta_{\alpha \beta})\,f_i\,,
\label{eq:rho_u_stress}
\end{equation}
where the lattice speed of sound $c_s^2=1/3$, the greek indices $\alpha$ and $\beta$ refer to tensor coordinates, i.e., $\bm{S}\!=\!\{S_{\alpha\beta}\}_{\alpha,\beta}$ for $\alpha,\beta\!\in\!\{x,\!y,\!z\}$.

\paragraph{High-order Distribution Reconstruction} 
With the moment-based representation~\cite{li2023high}, only macroscopic quantities $\mathcal{M}$ are stored in memory, achieving a balance between accuracy and computational efficiency. Distribution functions $f_i$ are then reconstructed on demand using a third-order closed-form Hermite expansion:
\begin{align} \label{eq:three_order_f}
f_i &=  \rho\, w_i \Biggl[ 1 + \frac{\bm{c}_i\cdot\bm{u}}{c_s^2} + \frac{\bm{H}^{[2]}(\bm{c}_i):\bm{S}}{2c_s^4}
    + \sum_{\alpha\beta\gamma}\smash{\frac{H^{[3]}_{\alpha\beta\gamma}(\bm{c}_i) \Gamma_{\alpha\beta\gamma}}{2c_s^6}}
\Biggr], \\ \nonumber 
&\Gamma_{\alpha\beta\gamma} = S_{\alpha\beta} u_\gamma 
  + S_{\alpha\gamma} u_\beta 
  + S_{\beta\gamma} u_\alpha 
  - 2\, u_\alpha u_\beta u_\gamma.\\[-6mm] \nonumber
\end{align}
Here, $w_i$ denote the lattice weights and $\bm{H}$ represents Hermite basis~\cite{Malaspinas:2015} and ${\alpha\beta\gamma} \!\in\! \{xxy,xyy,xxz,xzz,yzz,yyz,xyz\}$ denotes the Cartesian coordinate indices.

\paragraph{Collision and Streaming}
After streaming the reconstructed distributions, we compute three intermediate moments, $\rho^*$, $\rho^*\bm{u}^*$, and $\rho^*\bm{S}^*$ at each lattice node using~\autoref{eq:rho_u_stress}. 
Moment-based collision is applied to update $\rho$, $\rho\bm{u}$, and $\rho\bm{S}$ for next time-step
at each node.

\subsection{Particle-laden Flow}

In particle-laden flows, the interleaved distributions of fluid and sediment particles preclude explicit tracking of solid boundaries. From the local-averaged Navier-Stokes equation~\cite{Gao2018Mixture}, the mass and momentum conservation equations of fluid are
\begin{align}
\frac{\partial (\bbzeta \bbrho^\text{f})}{\partial \bbt} +\nabla \cdot(\bbzeta \bbrho^\text{f}\bbu) &=0 \label{eq:fluid} \\
\frac{\partial (\bbzeta \bbrho^\text{f}\bbu)}{\partial \bbt}+\nabla \cdot(\bbzeta \bbrho^\text{f}\bbu\otimes\bbu) &=-\bbzeta \nabla \bbp+\bbzeta \bbrho^\text{f} \bbg+\bbf^\text{f}
\label{eq:fluid_mom} 
\end{align}
where $\bbzeta$, $\bbrho^\text{f}$, $\bbu$, $\bbg$, $\bbp$, and $\bbf^\text{f}$ represent the fluid volume fraction, density, velocity, pressure, gravitational constant, pressure, and fluid drag force density from the sediment to the fluid, respectively and  $\otimes$ indicates the outer product;
Similarly, the sediment equations read:
\begin{align}
\frac{\partial (\bbeta \bbrho^\text{s})}{\partial \bbt} +\nabla \cdot(\bbeta \bbrho^\text{s}\bbv) &=0 \label{eq:sediment} \\
\frac{\partial (\bbeta \bbrho^\text{s}\bbv)}{\partial \bbt} +\nabla \cdot(\bbeta \bbrho^\text{s}\bbv\otimes \bbv) &=\bbeta \nabla \cdot \bbsigma^\text{s}+\bbeta \bbrho^\text{s}\bbg+\bbf^\text{s}
\label{eq:sediment_mom} 
\end{align}
where $\bbeta$, $\bbrho^\text{s}$, $\bbv$, and $\bbsigma^\text{s}$ represent the sediment volume fraction, density, velocity, and the Cauchy stress describing the mechanical responses inside solid clumps, respectively. 

From mixture theory, the volume fraction of fluid and sediment obeys $\bbzeta + \bbeta = 1$. In the case of debris in the air without liquid, $\bbzeta = 0$ and $\bbeta = 1$. Note that $\bbf^\text{f}$ and $\bbf^\text{s}$ denote the interaction drag force of the sediment on the fluid and the force of the fluid
on the sediment. As such, they obey $\bbf^\text{f} = -\bbf^\text{s}$. From the assumption that $\epsilon + \delta = 1$, combining~\autoref{eq:fluid} and \ref{eq:sediment} leads to
\begin{align}
\nabla \cdot (\bbzeta\bbu+\bbeta\bbv) = 0
\end{align}

\section{Particle-laden Flow in LBM}

In kinetic theory, the viscous incompressible Navier–Stokes equations can be recovered from the discrete Boltzmann equation with a BGK collision operator~\cite{Harris2012intro}. Accordingly, we first reformulate the locally averaged fluid equation (\autoref{eq:fluid}) into the standard Navier–Stokes form and then convert it into the corresponding Lattice Boltzmann formulation.
The fluid and sediment momentum equations (\autoref{eq:fluid_mom} and~\autoref{eq:sediment_mom}) can be summed to yield
\begin{align}
\bbrho^\text{f} \frac{\partial(\epsilon\bbu + \eta\bbv)}{\partial \bbt}  & + \bbrho^\text{f} \nabla \cdot(\epsilon\bbu\otimes\bbu + \eta\bbv\otimes\bbv) = 
\label{eq:combined} \\
& -\epsilon\nabla\bbp + \frac{\bbrho^\text{f}}{\bbrho^\text{s}} \eta \nabla \cdot \bbsigma^\text{s} + (\epsilon+\eta)\bbrho^\text{f} \bbg + \bbf^\text{f} + \frac{\bbrho^\text{f}}{\bbrho^\text{s}}\bbf^\text{s} \notag 
\end{align}
We define the relative velocity $\bbq=\bbv-\bbw$ and the mixture-averaged velocity $\bbw=\epsilon\bbu+\eta\bbv$.
Since $\epsilon+\eta=1$, the fluid velocity can be expressed as
\begin{equation}
\bbu = \bbw-\frac{\eta}{\epsilon}\bbq \;.
\end{equation}
The convective term can then be expanded as
\begin{align}
\epsilon \bbu\otimes\bbu + \eta \bbv\otimes\bbv = \bbw\otimes\bbw+\frac{\eta}{\epsilon} \bbq\otimes\bbq \;.
\end{align} 
Substituting these expressions into the combined momentum equation (\autoref{eq:combined}) yields
\begin{align}
\bbrho^\text{f}\frac{\partial \bbw}{\partial \bbt} +\bbrho^\text{f}\nabla \cdot(\bbw\otimes\bbw) &-\frac{\eta}{\epsilon}\bbrho^\text{f}\nabla \cdot(\bbq\otimes\bbq) =\\
&-\epsilon\nabla \bbp+\frac{\bbrho^\text{f}}{\bbrho^\text{s}} \eta\nabla \cdot \bbsigma^\text{s}  +\bbrho^\text{f}\bbg+\bbf^\text{f} + \frac{\bbrho^\text{f}}{\bbrho^\text{s}}\bbf^\text{s}\notag
\end{align}
Assuming that the fluid density is much smaller than the sediment density ($\bbrho^\text{f} \ll \bbrho^\text{s}$), and the drag force is sufficiently strong to keep the relative velocity $\bbq$ small, higher-order terms involving $\bbq$ and sediment stress can be neglected. Under these assumptions, we obtain the momentum equation governing mixture-averaged velocity:
\begin{equation}
\bbrho^\text{f}\frac{\partial \bbw}{\partial \bbt} +\bbrho^\text{f}\nabla \cdot(\bbw\otimes\bbw)=-\epsilon\nabla \bbp+\bbrho^\text{f}\bbg+\bbf^\text{f}
\label{eq:mom}
\end{equation}
The corresponding continuity equation reduces to the divergence-free condition $\nabla\cdot\bbw=0$, which is independent of the sediment motion. Finally, \autoref{eq:mom} can be equivalently rewritten as
\begin{equation}\label{eq:reformulated_momentum}
\bbrho^\text{f}\frac{\partial \bbw}{\partial \bbt} +\bbrho^\text{f}\nabla \cdot(\bbw\otimes\bbw)=-\nabla (\epsilon \bbp)+\bbp\nabla\epsilon+\bbrho^\text{f}\bbg+\bbf^\text{f}
\end{equation}

In the LBM formulation, the standard pressure is obtained from the Chapman–Enskog expansion~\cite{Harris2012intro} as $\epsilon \bbp=  \rho c_s^2$. Under this formulation, the additional term $\mathbf{p}\nabla \epsilon$ can be interpreted as an external force term 
$C\rho c_s^2\nabla\epsilon/\epsilon$, 
where $C = \frac{\bbrho \bbu_\text{ref}^2}{u_\text{ref}^2 \Delta \bbx}$ is the conversion factor from normalized lattice units to physical units~\cite{Wei2020}.
In practice, however, we observe that this additional force term introduces numerical instabilities in the LBM solver, leading to noticeable violations of the incompressibility condition during the simulation.
{To alleviate this issue, we adopt a modified pressure formulation, $\epsilon \bbp=(\rho-\rho_0) c_s^2$ where $\rho_0 = 1$ is the reference density in lattice units. }
This modification effectively removes the constant background pressure and significantly improves numerical stability while preserving the incompressible flow behavior. Hence, we can get external force $\FF$ in Lattice space as~\cite{Zhang2014}:
\begin{equation}
\FF=
  \frac{\rho-\rho_0}{\epsilon} \nabla\epsilon+
C^{-1} (\bbrho^\text{f}\bbg+\bbf^\text{f}) \; .
\end{equation}


To account for mixture effects in the BGK-based Lattice Boltzmann formulation, the effective viscosity in the collision term must be modified accordingly. As shown by \citet{Tsigginos2019pre}, this can be achieved by scaling the relaxation time with the mixture fraction $\epsilon$, leading to the following evolution equation:
\begin{equation}
\frac{\partial f_i}{\partial t} +\cc_i\cdot \nabla f_i=-\frac{f_i-f_i^\text{eq}}{\epsilon\tau}+F_i
\end{equation}

\section{Multi-level HOME-LBM Solver}
\label{sec:mlbm}

This section presents an adaptive block-based multi-level HOME-LBM solver, beginning with the multi-level block structure and the rescaling of moment-based physical parameters across levels. We then present the core multi-level LBM computation scheme. Finally, we discuss how the proposed solver is coupled with MPM to support complex physical phenomena.

\subsection{Multi-level Block Structure}

\begin{figure}[ht!]
\centering
\includegraphics[trim = 30 0 30 0, clip, width=.495\linewidth]{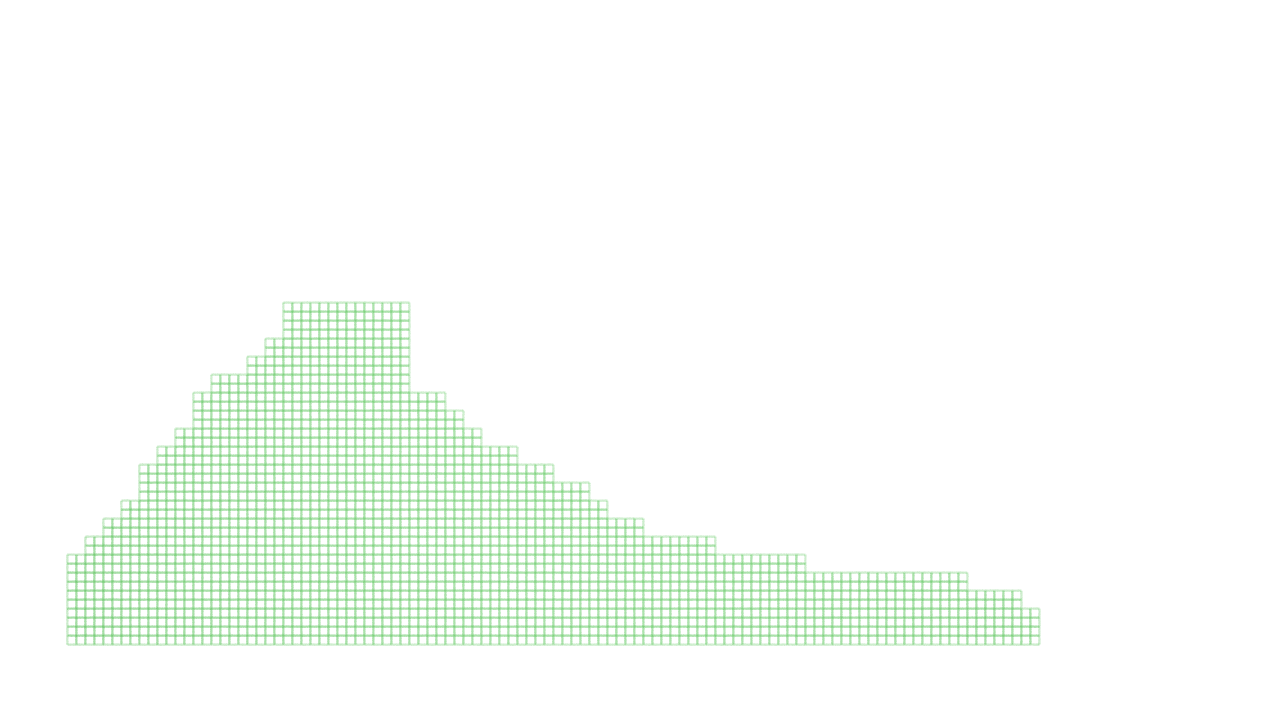}
\includegraphics[trim = 30 0 30 0, clip, width=.495\linewidth]{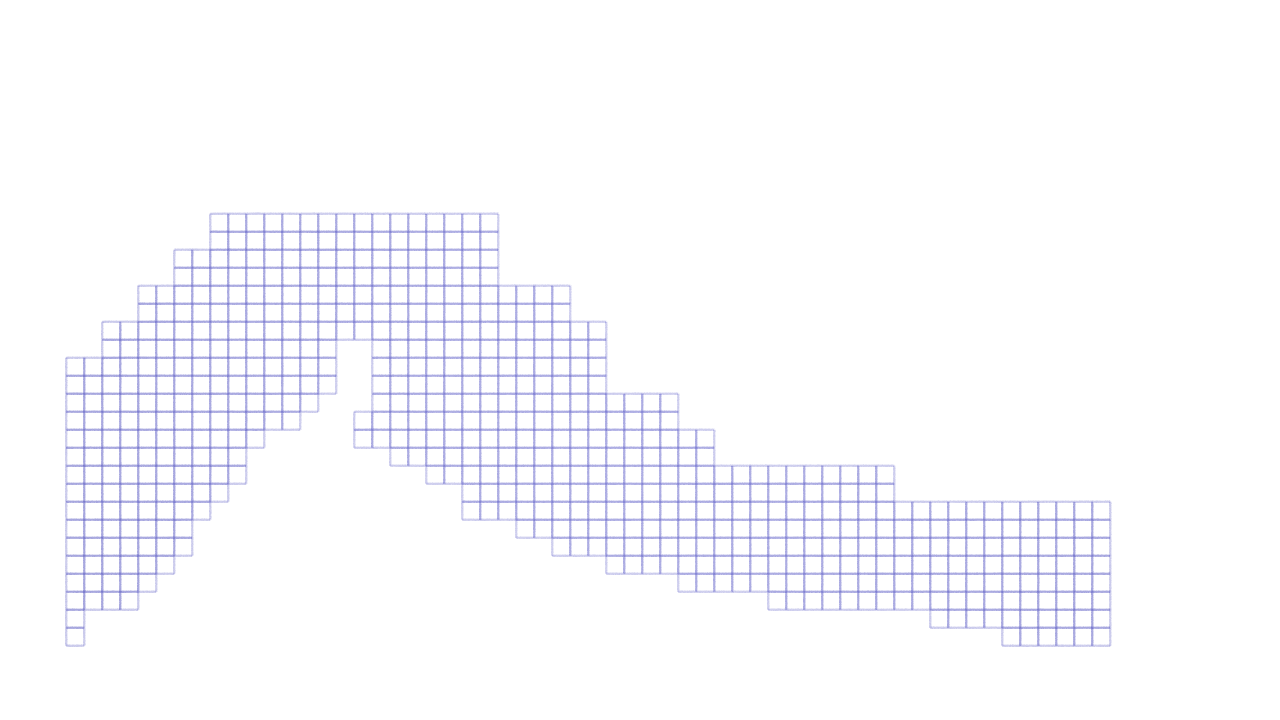}
\put(-150,50){\small $l=0$ }
\put(-30,50){\small $l=1$ }\\
\includegraphics[trim = 30 0 30 0, clip, width=.495\linewidth]{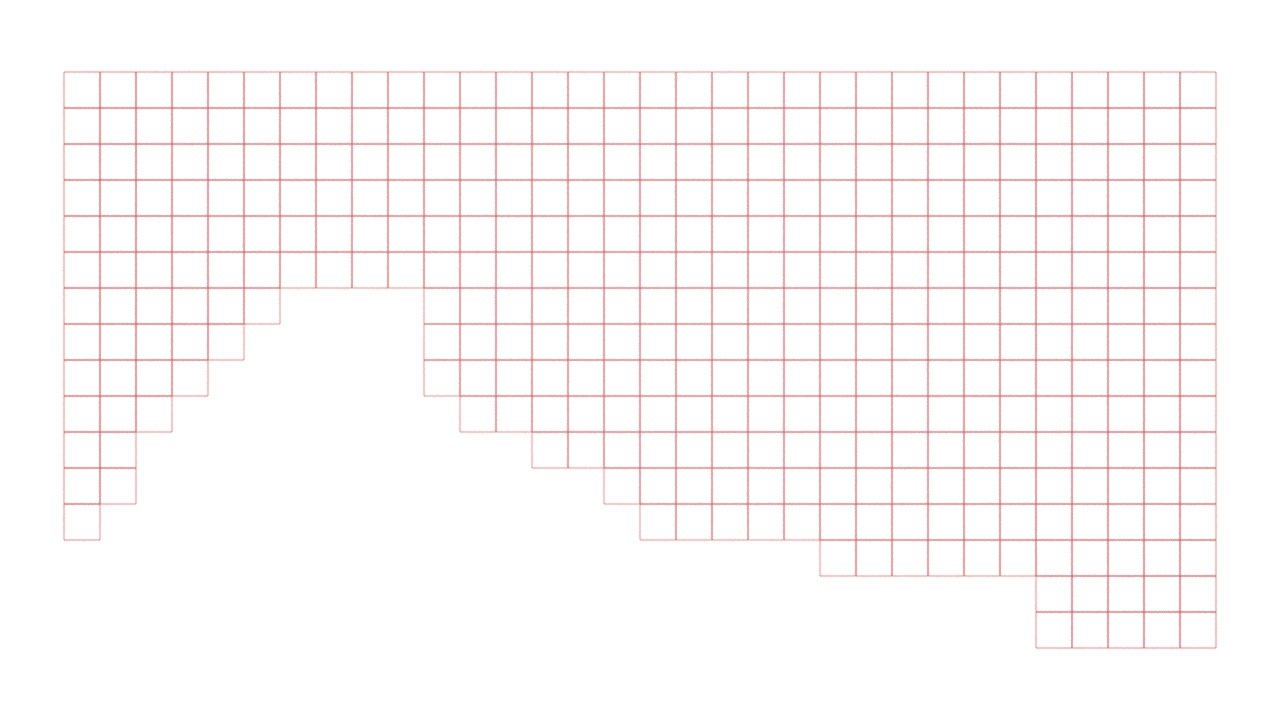}
\includegraphics[trim = 30 0 30 0, clip, width=.495\linewidth]{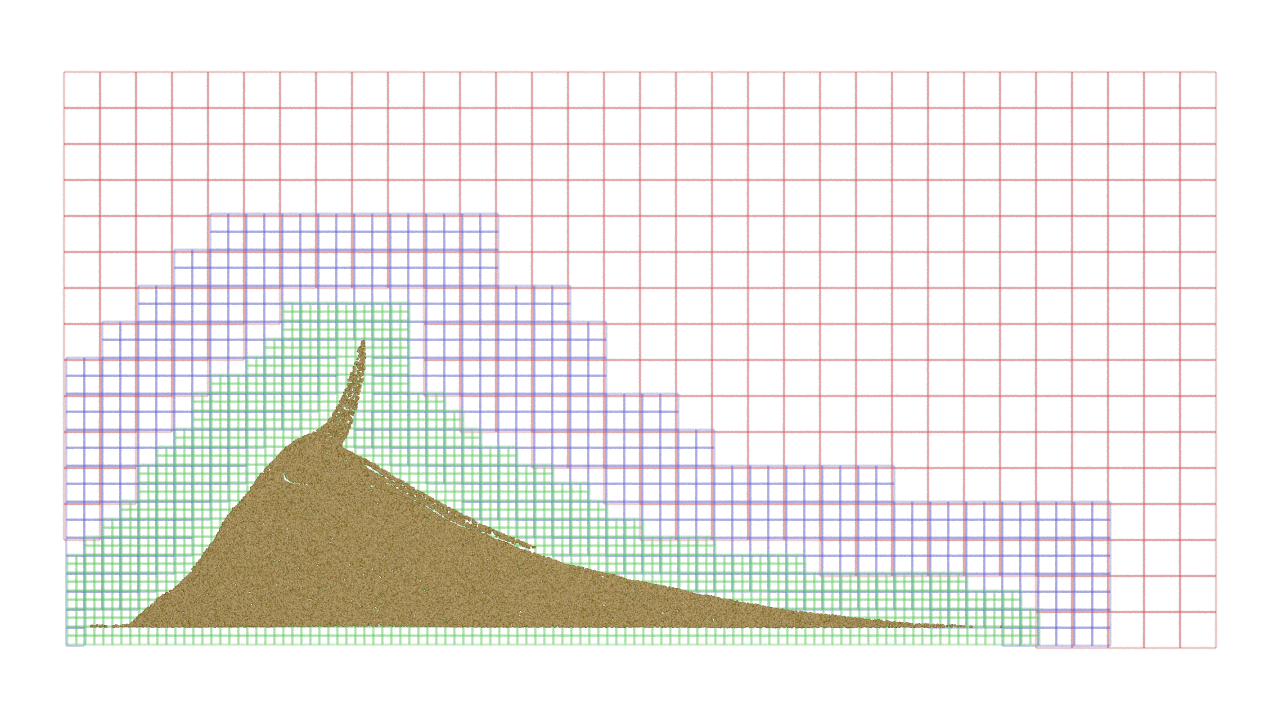}
\put(-150,50){\small $l=2$ }
\put(-30,50){ \small All }
\vspace{-0.5em}
\caption{Multi-level block structure in 2D.
Yellow denotes sand particles. Red, blue, and green denote different levels.}
\label{fig:lattice_structure_bird}
\Description{}
\end{figure}

In our multi-level HOME-LBM framework, the block hierarchy organizes multiple blocks at different spatial and temporal resolutions to capture flow physics across scales. The computational domain is decomposed into a finite set of lattice levels (\autoref{fig:lattice_structure_bird}), where level $l=0$ denotes the finest resolution, and each level $l$ consists of possibly disconnected uniform Cartesian grids with spacing $\Delta x_l = 2^{l}\Delta x_0$. Each node at level $l$ can be subdivided into $2^d$ nodes at level $l-1$, with $d$ being the spatial dimension. Following~\cite{Lagrava2012lbm}, coarse cells are removed when fully covered by fine cells, except in the interface regions. Note that we keep the finer grid with one block shift, as shown in~\autoref{fig:lattice_structure_bird}, to align grids of different levels at certain grid nodes.


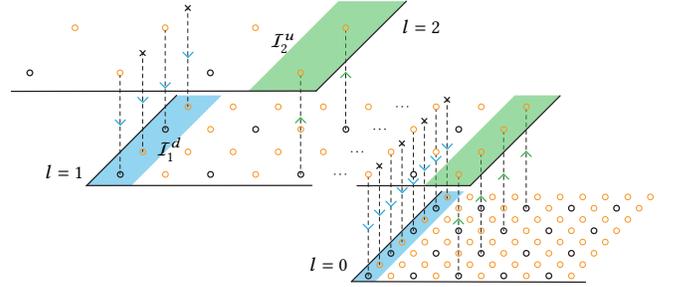
\begin{figure}[ht!]
\centering
\includegraphics[trim = 0 0 0 0, clip, width=\linewidth]{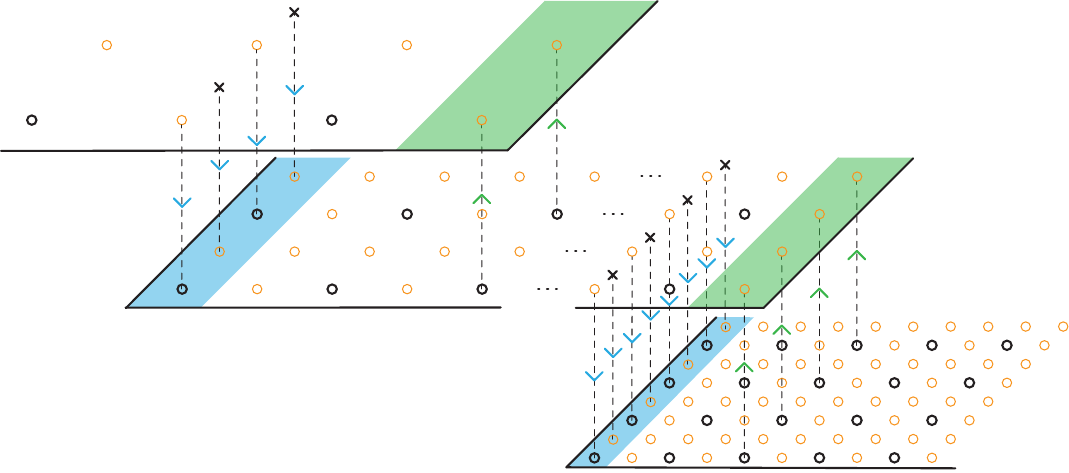}
\put(-130,5){\footnotesize $l=0$}
\put(-230,40){\footnotesize $l=1$}
\put(-95,95){\footnotesize $l=2$}
\put(-188,49){\footnotesize $\mathcal{I}^d_{1}$}
\put(-145,90){\footnotesize $\mathcal{I}^u_{2}$}
\caption{Blue/green arrows and areas denote downward/upward transfer and interface areas $\mathcal{I}^d_{\bullet}$/$\mathcal{I}^u_{\bullet}$, respectively. Black circles indicate corresponding children who transfer macroscopic quantities to their parents during the upward transfer.}
\label{fig:lattice_structure}
\Description{}
\end{figure}

We also explicitly construct a parent–child mapping across the hierarchy (\autoref{fig:lattice_structure}), where each fine node has a unique parent at the next coarser level, and the coarse nodes may own multiple fine children that transfer macroscopic quantities, including density, velocity, and the second-order moment tensor, during simulation. At interfaces between adjacent levels, both levels maintain overlapping interface areas to facilitate data transfer across level boundaries. Macroscopic quantity exchange occurs only at boundaries, not across the full overlap region.


\subsection{Physical Rescaling}

When transferred between hierarchy levels, macroscopic quantities must be adjusted consistently to preserve physical behavior. Following the order-of-magnitude analysis of~\citet{Lagrava2012lbm}, the kinematic viscosity $\nu$ and relaxation time $\tau$ rescale between levels $l$ and $l+k$ as
\begin{equation}
\nu_{l+k}=\frac{\nu_l}{2^k}\text{,}\quad
\tau_{l+k}=\frac{1}{2^k}\tau_{l}+\frac{2^{k}-1}{2^{k+1}}
\end{equation}

For the moment variables, the density $\rho$ and velocity $\uu$ remain invariant under proportional rescaling of space and time~\cite{li2018}. The discrete local equilibrium distribution $\bm{f}^\text{eq}$ is constructed using a second-order approximation of the Maxwell–Boltzmann distribution~\cite{Vardhan2019lbm}:
\begin{equation}
f^\text{eq}_i= \rho w_i \left( 1+ \frac{\bm{c}_i \bm{u}}{c^2_s} + 
\frac{(\bm{c}_i \bm{u})^2}{2c^4_s} - \frac{\bm{u}^2}{2c^2_s}\right) \; ,
\label{eq:lbe_feq}
\end{equation}
which depends only on $\rho$ and $\uu$ and therefore remains unchanged across levels. From the equilibrium distribution, the equilibrium and non-equilibrium components of the second-order moment tensor $\SS$ are given by
\begin{align}
S_{\alpha\beta}^\text{eq} &=\frac{1}{\rho}\sum_i\left(c_{i\alpha}c_{i\beta}-c_s^2\delta_{\alpha\beta}\right)f_i^{\text{eq}} \label{eq:Seq} \\
S_{\alpha\beta}^\text{neq} &=\frac{1}{\rho}\sum_i c_{i\alpha}c_{i\beta}f_i^{\text{neq}} \label{eq:Sneq} \;.
\end{align}
respectively, since $\sum_i f_i^\text{neq}=\rho-\sum_i f_i^\text{eq}=0$. Combining \eqref{eq:lbe_feq} and \eqref{eq:Seq},  we get $S^\text{eq}=S^\text{eq}(\rho,\uu)$ so that it is also invariant across levels. By Chapman–Enskog analysis~\cite{Harris2012intro}, the non-equilibrium component satisfies
\begin{equation}
S_{\alpha\beta}^\text{neq}\propto -c_s^2\tau\left(\partial_\alpha u_\beta+\partial_\beta u_\alpha\right) \; ,
\end{equation}
implying that $\SS^\text{neq}$ rescales across levels according to
\begin{equation}
\SS^\text{neq}_{l+1}=\frac{\tau_{l+1}}{2\tau_l}\SS^\text{neq}_l \;.
\end{equation}
Finally, since $\SS=\SS^\text{eq}+\SS^\text{neq}$, we obtain the rescaling rule for the full second-order moment tensor:
\begin{equation}\label{eq:upload_S}
\SS_{l+1}=\frac{\tau_{l+1}}{2\tau_l}\SS_l+(1-\frac{\tau_{l+1}}{2\tau_l})\SS^\text{eq}
\end{equation}

\subsection{Multi-level LBM Advance Scheme}

Based on the traditional multi-level LBM framework~\cite{Lagrava2012lbm}, we develop a moment-based multi-level LBM solver, summarized in~\autoref{alg:mlbm}. 
To keep the lattice velocity constant across levels, each level advances with its own time step, with the time step proportional to the spatial resolution, i.e., $\Delta t_{l+1}=2\Delta t_l$. As a result, finer levels perform multiple smaller steps per coarse-level step, resulting in a sub-cycling scheme in which streaming and collision are executed recursively across levels. After one streaming–collision step at level $l$, two successive steps are performed at level $l-1$, referred to as step~1 and step~2 (\autoref{fig:pipeline}).

Before each streaming–collision step at level $l$, macroscopic quantities are gathered from the coarser level $l+1$ at the downward interface region $\mathcal{I}_{l}^d$. To limit inter-level data movement, the corresponding upward transfer is performed only after the second fine-level step to the upward interface region $\mathcal{I}_{l+1}^u$ at level $l+1$, returning macroscopic quantities to the coarser level. As illustrated in~\autoref{fig:lattice_structure}, we follow~\citet{Lagrava2012lbm} to use a two-block separation, since insufficient separation between downward and upward interface regions will lead to bidirectional dependency loops. 



 

\begin{algorithm}[ht]
\caption{Multi-level LBM solver}
\label{alg:mlbm}
\begin{algorithmic}[1]
    \Function{AdvanceLBM}{level $l$, step $s$}
        \If{$l<L-1$}
            \State \Call{DownwardTransfer}{$l+1$, $s$}
        \EndIf
        \State \Call{StreamingCollision}{$l$}
        \If{$l<L-1$ and $s=2$}
            \State \Call{UpwardTransfer}{$l$}
        \EndIf
        \If{$l>0$}
            \State \Call{AdvanceLBM}{$l-1$, 1} \Comment{step 1}
            \State \Call{AdvanceLBM}{$l-1$, 2} \Comment{step 2}
        \EndIf
    \EndFunction

\end{algorithmic}
\end{algorithm}



\paragraph{Downward Transfer}
For nodes in the downward interface region $\mathcal{I}_l^d$ at level $l$, a one-to-one correspondence with nodes at level $l+1$ does not always exist. We therefore identify the nearest $2^D$ (D is grid dimension) neighboring nodes at level $l+1$ and interpolate the macroscopic quantities as 
\begin{align}
\rho_{l,i}&=\sum_j w_{ij}\rho_{l+1,j} ,  \; 
\uu_{l,i}=\sum_j w_{ij}\uu_{l+1,j} , \label{eq:downward_rescale} \\
\SS^\prime&=\frac{2\tau_{l+1}}{\tau_{l}}\sum_j w_{ij}\SS^{\text{neq}}_{l+1,j} + \SS^\text{eq} \; . \notag
\end{align}
In addition, the second streaming–collision step at level $l$ occurs at the temporal midpoint of the advance at level $l+1$. Beyond spatial interpolation, we perform temporal interpolation of macroscopic quantities between $\mathcal{M}^{l+1,t}$ and $\mathcal{M}^{l+1,t+2\Delta t_l}$ to obtain $\mathcal{M}^{l,t+\Delta t_l}$. 



\paragraph{Upward Transfer}
For each node in the upward interface region $\mathcal{I}_{l+1}^u$ at level $l+1$, we fetch and rescale the macroscopic quantities from its corresponding node at level $l$ according to~\autoref{eq:upload_S}.

\subsection{Coupling with Material Point Method (MPM)}

Our framework employs MPM to simulate granular materials such as snow and sand. readers are referred to prior work~\cite{Klar2016Sand,Stomakhin2013} for detailed formulations.

In particle-laden flows, the most significant interactions occur in the vicinity of sediment particles. To ensure accurate two-way coupling, we enforce that all MPM particles reside within LBM nodes at the finest level ($l=0$). Following~\citet{Wang2025cirrus}, we define a level-selection function which adaptively refines the lattice only where particle–fluid interactions are present.
\begin{equation}
\Phi(\mathcal{T})=
\begin{cases}
0, &\text{if $\mathcal{T}$ contains a MPM particle},\\
L-1, &\text{otherwise}
\end{cases}
\end{equation}
which adaptively refines the lattice only where particle–fluid interactions are present.

\autoref{fig:pipeline} illustrates a complete LBM–MPM advance. After the LBM streaming step at level $l=0$, we compute the fluid and sediment volume fractions and the corresponding volumetric drag forces during the MPM grid update. These quantities are then fed back into the LBM collision step. Drag is modeled using the Di Felice formulation~\cite{Gao2018Mixture} to couple fluid and sediment within each grid cell. In practice, the MPM time step is typically larger than the LBM time step at level $l=0$, and the MPM update is therefore performed once every several (e.g., 3-5) LBM steps.

Compared with a full-resolution LBM solver, our adaptive block-based multi-level approach takes the same number of integration steps at the finest level.
Crucially, fine-grid computations are performed only in blocks where high resolution is needed; the rest of the domain advances on coarser blocks, avoiding the cost of maintaining fine grids everywhere. As a result, the overall computational expense is substantially reduced.


\section{GPU Implementations}

This section describes the GPU memory layout of our multi-level data structure, recursive ping-pong buffering, and the dynamic update of the structure in response to MPM particle motion.

\subsection{Memory Layout}
All LBM macroscopic quantities are stored in a tile-based, octree-like hierarchy, where each tile $\mathcal{T}$ contains $4$ cells per axis ($4\times4\times4$ in 3D), and memory is allocated and accessed at tile granularity. The MPM sparse-grid topology reuses the finest-level LBM tiles. At each level, we maintain a table of active tiles with a spatial hash for efficient lookup by level and position. For stability, we enforce a two-tile overlap in each direction, ensuring that for non-overlapping tiles all first- and second-ring neighbors reside at the same level.
To reduce memory usage, we store only \leaf~and \border~tiles: leaf tiles have no parent or child across levels, while border tiles overlap with adjacent levels; internal tiles that do not participate in streaming transmit no data to external nodes and are therefore excluded from the fluid solve.
This significantly lowers memory consumption for sparse domains. All per-tile data are stored in a structure-of-arrays (SoA) layout to ensure coalesced memory access~\cite{Chen2022}.


\subsection{Recursive Ping-Pong Buffering}

In our multi-level LBM solver, temporal interpolation at level $l$ requires access to both the previous and updated macroscopic quantities at the coarser level $l+1$. Naively maintaining full buffers at every level results in a worst-case memory footprint that scales as $2^L$, where $L$ denotes the number of hierarchy levels. To avoid this exponential growth, we introduce a recursive ping-pong buffering scheme that maps the current and next states of all levels onto only two global data trees.
At a given bounce, the current states of even-numbered levels (e.g., $l=0,2,4,\dots$) are stored in tree~A and those of odd-numbered levels (e.g., $l=1,3,\dots$) in tree~B, with the next states assigned to the opposite trees. The assignments are swapped at the next bounce. As a result, the multi-level advance only needs to track the bounce index to determine whether each level reads from or writes to tree~A or tree~B. This design reduces the memory requirement to two copies of the macroscopic field buffers across all levels, independent of hierarchy depth, rather than $2^L$ buffers.


\subsection{Dynamic Grid Update}
After each MPM iteration, we update both the MPM and LBM grids to reflect particle motion. In particular, a tile $\mathcal{T}$ at level $l$ is marked for refinement if $l > \Phi(\mathcal{T})$ or if any of its non-border neighboring tiles is marked for refinement.
We propose a GPU-based algorithm to efficiently maintain the dynamic multi-level LBM grid. Each tile carries one of three flags, \refine, \coarsen, or \delete, indicating that the tile should be refined, coarsened, or removed, respectively. Following each MPM iteration, the grid update proceeds in two stages: bottom-up refinement and top-down coarsening, ensuring hierarchy consistency while minimizing unnecessary grid updates.

\paragraph{Bottom-up Refinement} First, all tiles whose current level exceeds the target refinement level prescribed by $\Phi(\mathcal{T})$ are marked as \delete. The hierarchy is then processed level by level, from coarse to fine. Each tile is marked as \refine~if either the tile itself or any of its neighbors $\mathcal{N}(\mathcal{T})$ is scheduled for \delete, thereby ensuring neighborhood consistency across refinement boundaries. \refine~is propagated upward by marking each refined tile's parent $\mathcal{P}(\mathcal{T})$ as \delete~and resetting sibling tiles $\mathcal{S}(\mathcal{T})$ without children to \leaf, thereby enforcing valid parent–child configurations. Tiles marked as \delete~are removed from the current level and replaced by their children $\mathcal{C}(\mathcal{T})$, which are created at the next finer level and marked as \leaf, while tiles marked \refine~but lacking children are subdivided and labeled as \border. For all newly created fine-level tiles, macroscopic moments are initialized by interpolating from their parents, using~\autoref{eq:downward_rescale}.

\paragraph{Top-down Coarsening} It reduces the block hierarchy by merging fine tiles into coarser blocks while preserving neighborhood and connectivity constraints. Initially, all tiles whose current level is finer than the target level $\Phi(\mathcal{T})$ are marked as \coarsen. The hierarchy is then processed from fine to coarse levels, where sibling tiles $\mathcal{S}$ are first grouped into blocks $\mathcal{B}$. A block is marked as \coarsen~if all tiles in $\mathcal{B}$ are flagged as \coarsen, have no children, and form a complete $2^d$ group, while $\mathcal{B}$ is marked as \delete~if all neighboring blocks are also eligible for \coarsen~or if any neighboring block is not topologically connected, ensuring spatial consistency across block boundaries. During \coarsen, blocks marked as \delete~are replaced by their parents at the coarser level, which is created if necessary and marked as a \leaf, and all children in the block are removed. Blocks marked as \coarsen~but lacking an existing parent generate a new parent tile, and both the parent and its children are marked as \border. Finally, for each newly created coarse tile, macroscopic moments are transferred upward from the corresponding children.

\section{Results}

We demonstrate the effectiveness of our adaptive LBM-MPM coupled solver through a series of experiments. All experiments conduct one MPM step every LBM step at the finest level. The performance is collected on a desktop with a GPU with 32GB VRAM and 21760 cores and an Intel Xeon w7-2595X processor.

\begin{figure*}[!htb]
   \begin{minipage}{0.48\textwidth}
     \centering
\includegraphics[trim = 0 0 0 0, clip, width=\linewidth]{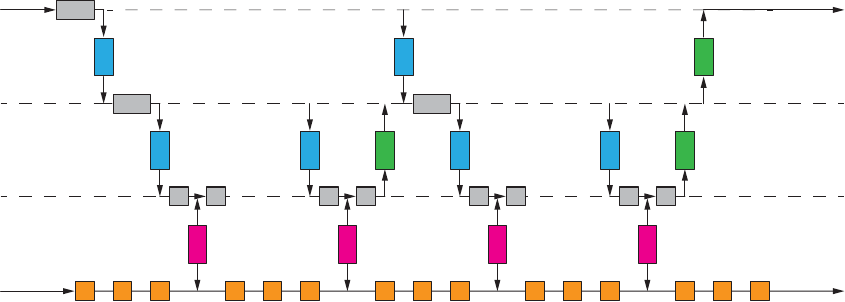}
\put(-240,-6){\scriptsize MPM}
\put(-240,22){\scriptsize LBM $l=0$}
\put(-240,49){\scriptsize $l=1$}
\put(-240,73){\scriptsize $l=2$}
\put(-224.5,82){\tiny SC}
\put(-208,55.5){\tiny SC}
\put(-193,28.5){\tiny S}
\put(-183,28.5){\tiny C}
\put(-202,-6){\tiny P2G}
\put(-212.5,-6){\tiny G2P}
\put(-220.5,-6){\tiny G}
\put(-183,15){\tiny Ex}
\caption{SC, S, and C denote the streaming–collision, streaming-only, and collision-only steps in the LBM advance, respectively. Ex indicates force exchange between LBM and MPM. G, P2G, and G2P denote grid update, particle-to-grid, and grid-to-particle operations in the MPM integration. Blue and green denote downward and upward transfers between LBM levels, respectively.}
\Description{}
\label{fig:pipeline}
   \end{minipage}\hfill
   \begin{minipage}{0.48\textwidth}
     \centering
\includegraphics[trim = 30 0 120 150, clip, width=\columnwidth]{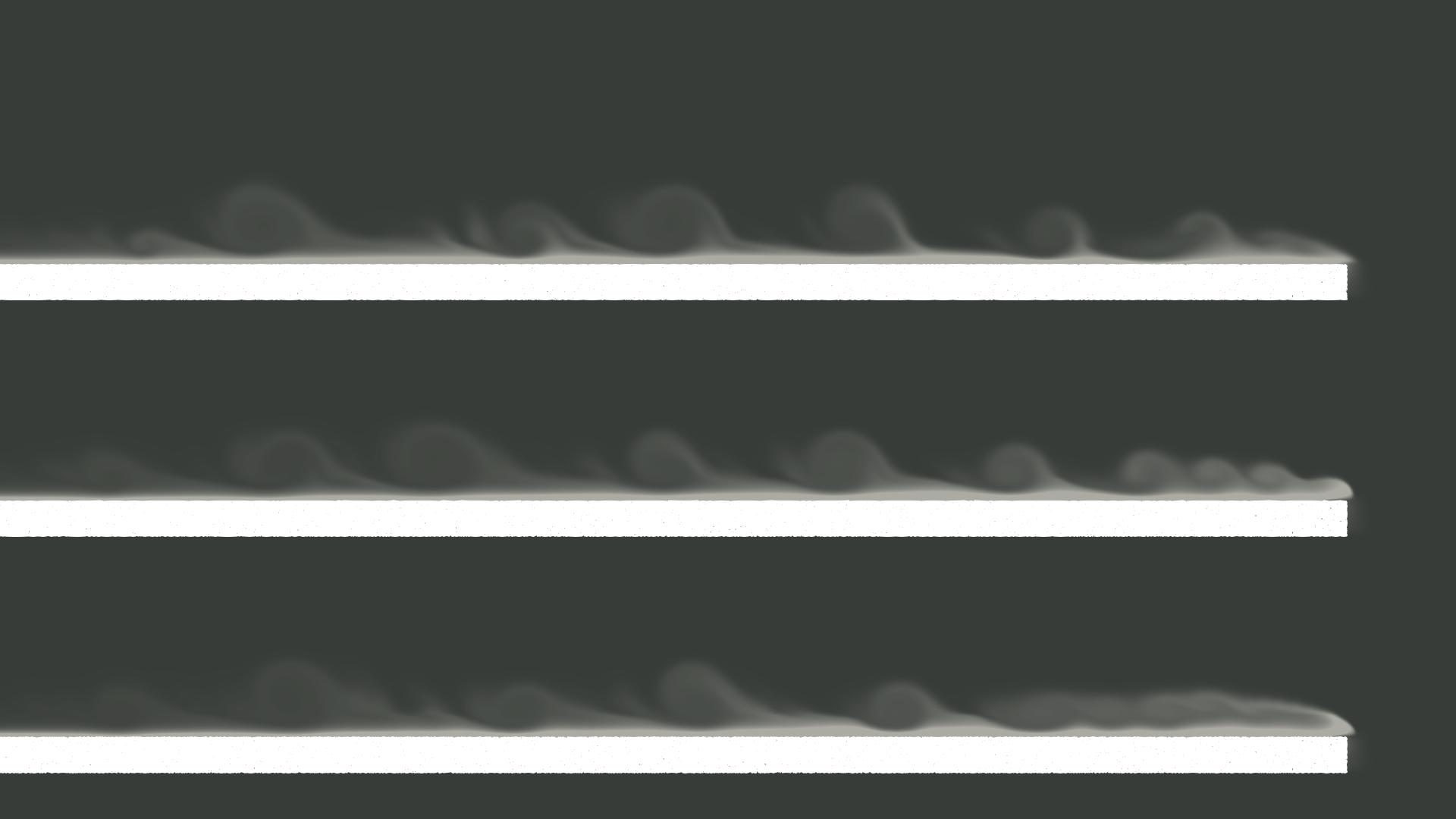}
\caption{\textbf{Powder Snow.} The powder snow cloud is generated with the advance of a bulk of snow pack. The entrainment rates from top to bottom are 1e-7, 4.5e-6, and 1e-6, respectively. A denser powder-snow cloud is less turbulent in our simulation.}
\label{fig:powder-snow}
   \end{minipage}
\end{figure*}

\paragraph{Powder snow} Based on the approximate powder-snow model~\cite{Nascimento2025}, we define the powder snow as a mixed fluid composed of air and snow in suspension. The volume fraction of the powder snow $\phi$ is maintained along with the fluid volume fraction
$\phi + \epsilon + \delta = 1$,
subjected to the transport-diffusion equation
$
\frac{\partial \phi}{\partial t} + \uu \cdot \nabla \phi = q_e - D \nabla^2 \phi \;.
$
To model powder snow coming from the entrainment process that occurs on the snow surface, we introduce the entrainment rate $q_e$ as a function of the shear stress $\tau$ at the snow region as
$
q_e = E (\vvv\cdot \boldsymbol{\tau})\cdot \vvv / \| \vvv \| $,
where $E$ is the entrainment coefficient, $\boldsymbol{\tau}$ is the stress tensor, and $\vvv$ is the snow velocity. In our simulation, we accumulate the powder snow growth on the dense grid in MPM steps and solve the transport-diffusion equation using a semi-Lagrangian RK3 advection and a forward-Euler diffusion. 
The resulting powder-snow simulation captures the realistic behavior of powder-snow avalanches as they flow down a slope. \autoref{fig:powder-snow} shows, as the entrainment rate increases, the velocity field becomes less turbulent, demonstrating our method's ability to track the complex flow mixture in powder snow dynamics.

\begin{figure*}[ht!]
\centering
\includegraphics[trim = 6cm 0 6cm 0, clip, width=0.33\linewidth]{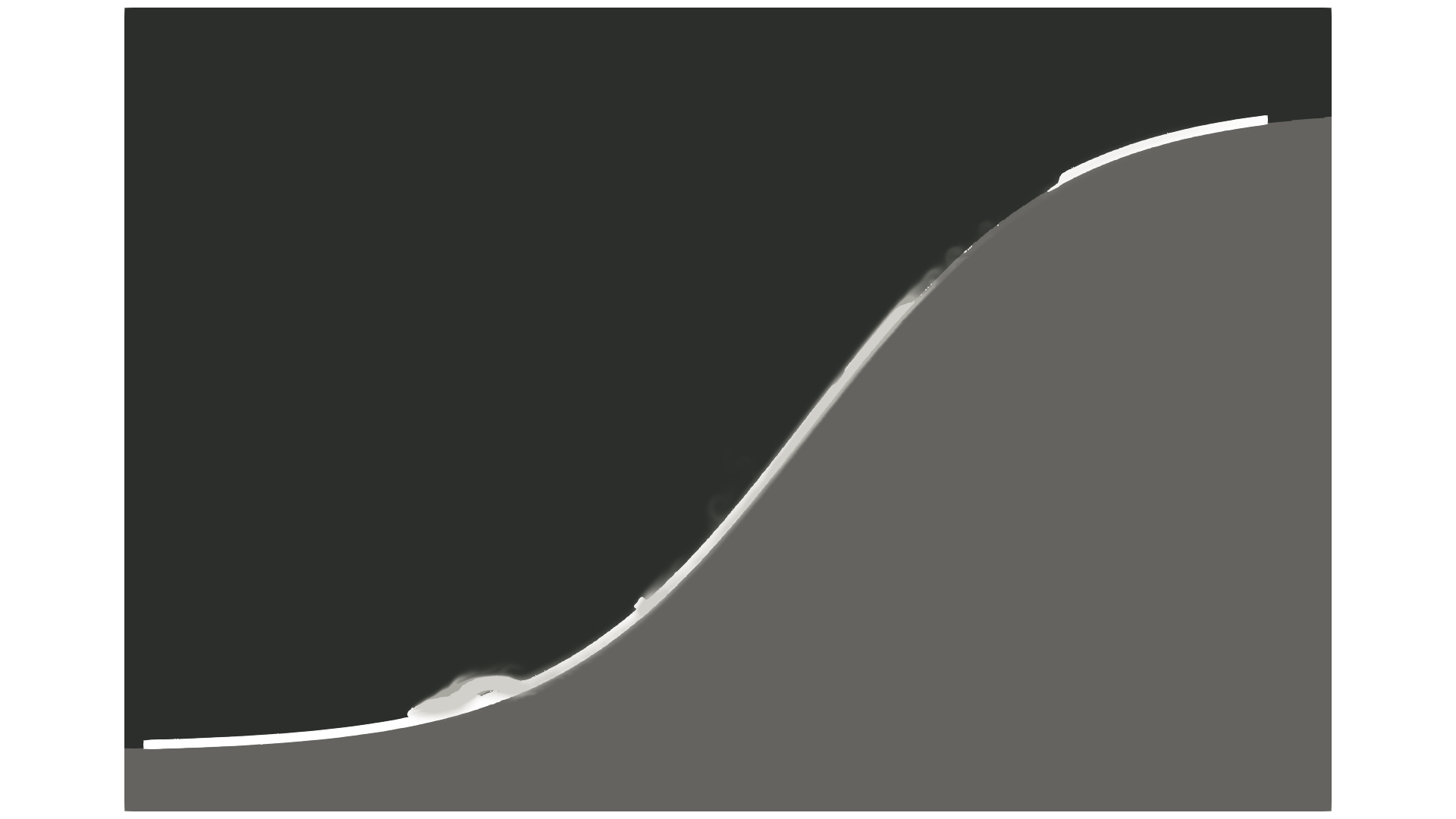}
\includegraphics[trim = 6cm 0 6cm 0, clip, width=0.33\linewidth]{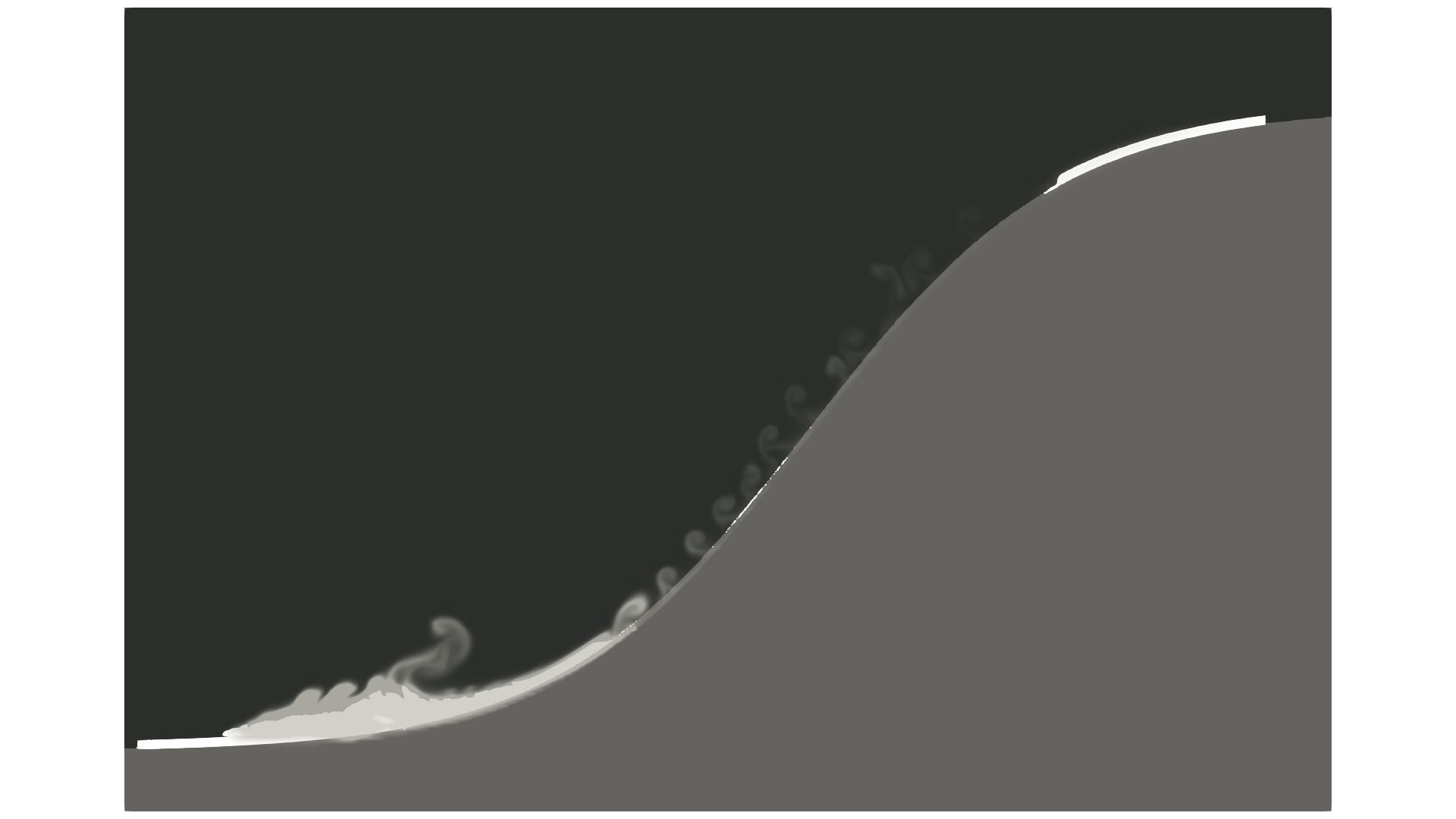}
\includegraphics[trim = 6cm 0 6cm 0, clip, width=0.33\linewidth]{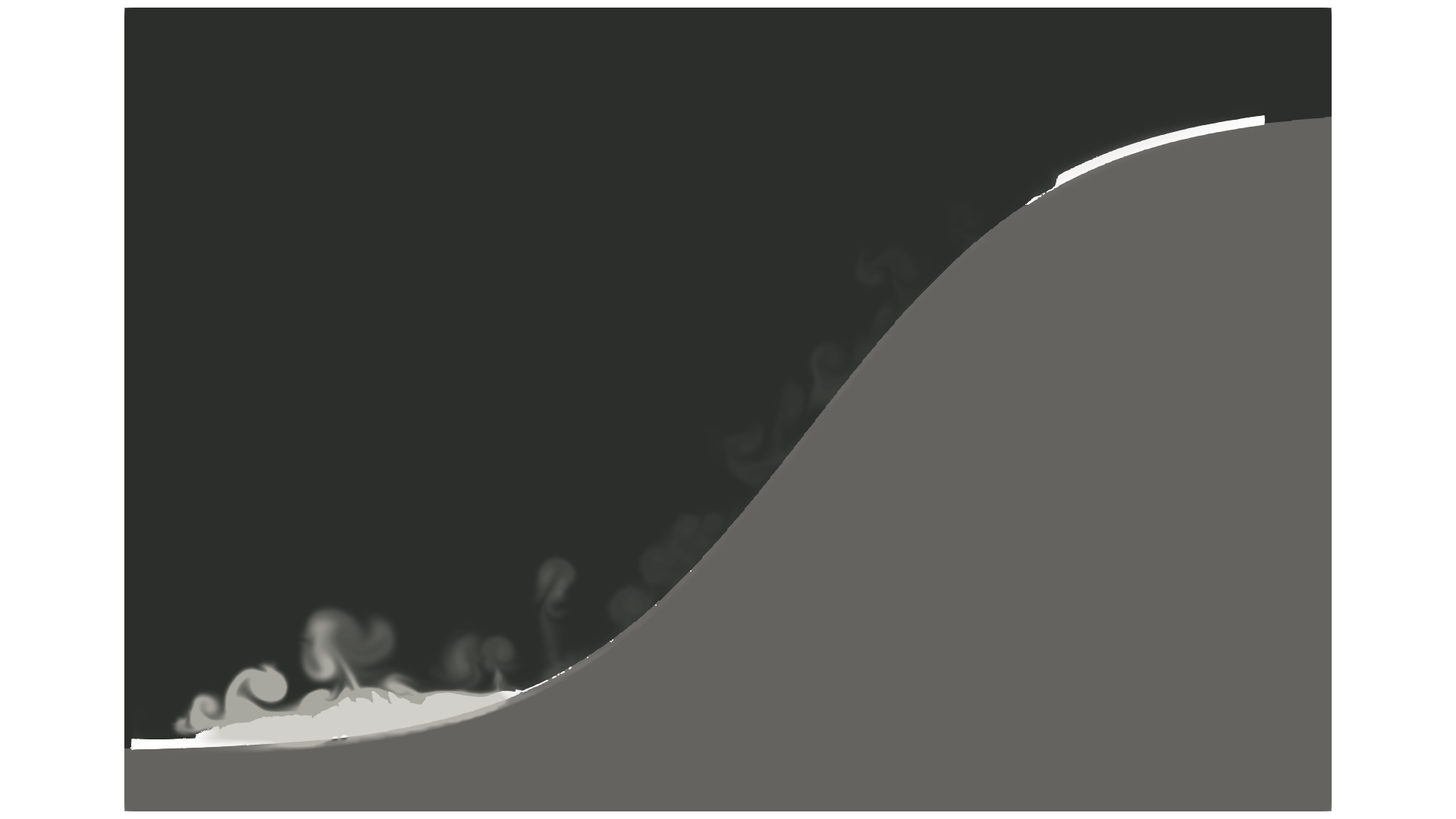}
\caption{\textbf{2D Avalanche on a slope.} A high-fidelity powder snow avalanche simulation descending a curved slope. Typical phenomena, including aerodynamic entrainment, vortex shedding, and the turbulent suspension cloud driven by Kelvin-Helmholtz instability, are reproduced.}
\label{fig:avalanche-2d}
\Description{}
\end{figure*}

\paragraph{Avalanche on a slope} 
Following~\citet{gaume2018}, we modify the Non-Associated Cam Clay (NACC)~\cite{wolper2019} with a soft layer to simulate the subtle crack and collapse behavior of snowpack. Specifically, the snow particles are set to softened state initially, and change according to the hardening law
\begin{equation}
    \frac{dq}{dt}=\begin{cases}
        -\alpha \dot{q}_0 & \text{if never } q=0 \;, \\
        \dot{q}_0 & \text{if once } q=0 \;.
    \end{cases}
\end{equation}
where $q$ is the hardening parameter, $\alpha$ is the soften coefficient, and $\dot{q}_0$ is the original hardening rate in NACC. At the same time, if we find $q=0$ on a particle, then the cohesive coefficient at the particle is also set to zero accordingly. This modification allows the snowpack to behave like a fluid after the crack, gradually hardening as it flows down the slope, as shown in~\autoref{fig:avalanche-2d}. 

\begin{figure*}[ht!]
\centering
\includegraphics[trim = 500 210 600 260, clip, width=0.33\linewidth]{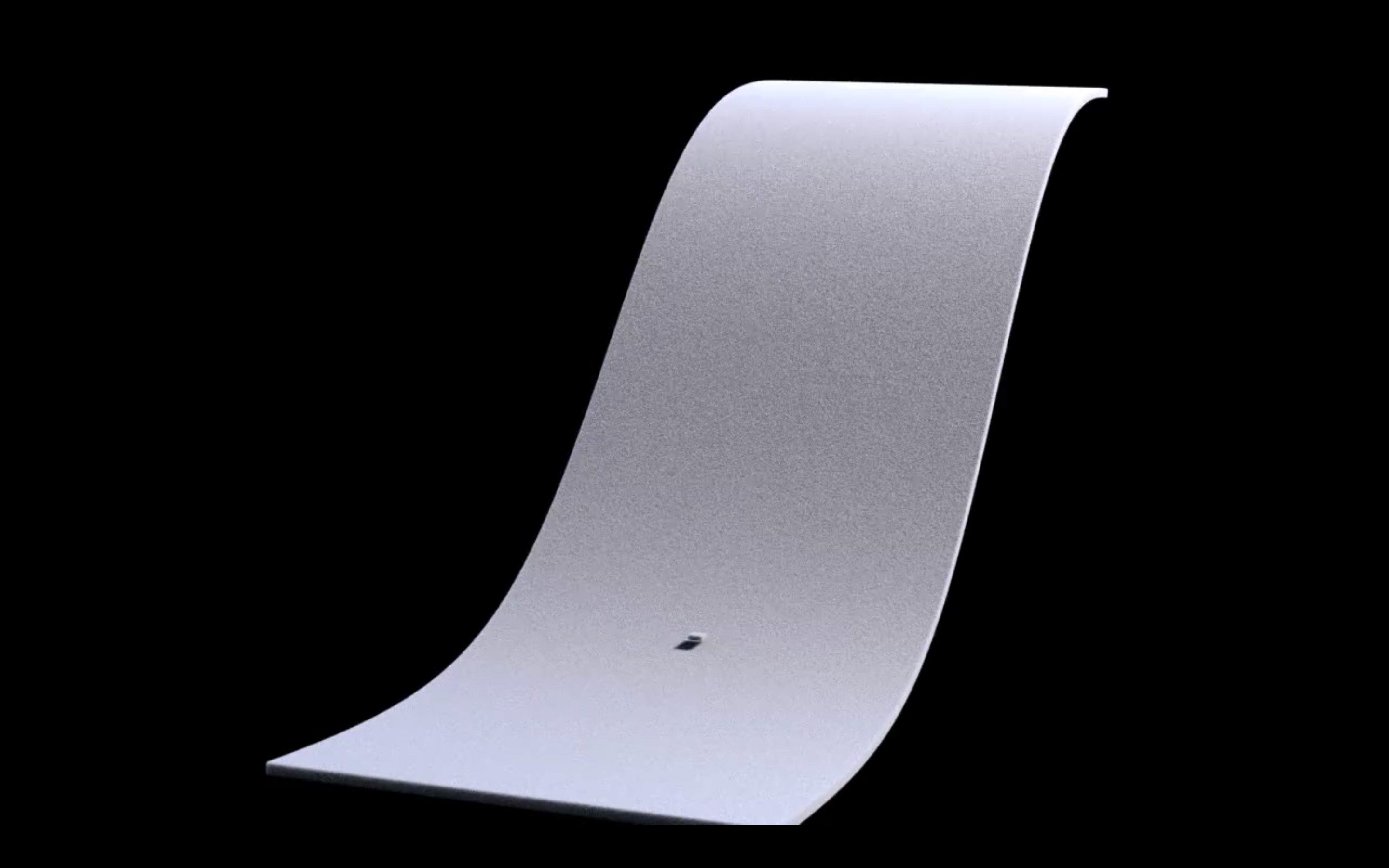}
\includegraphics[trim = 500 210 600 260, clip, width=0.33\linewidth]{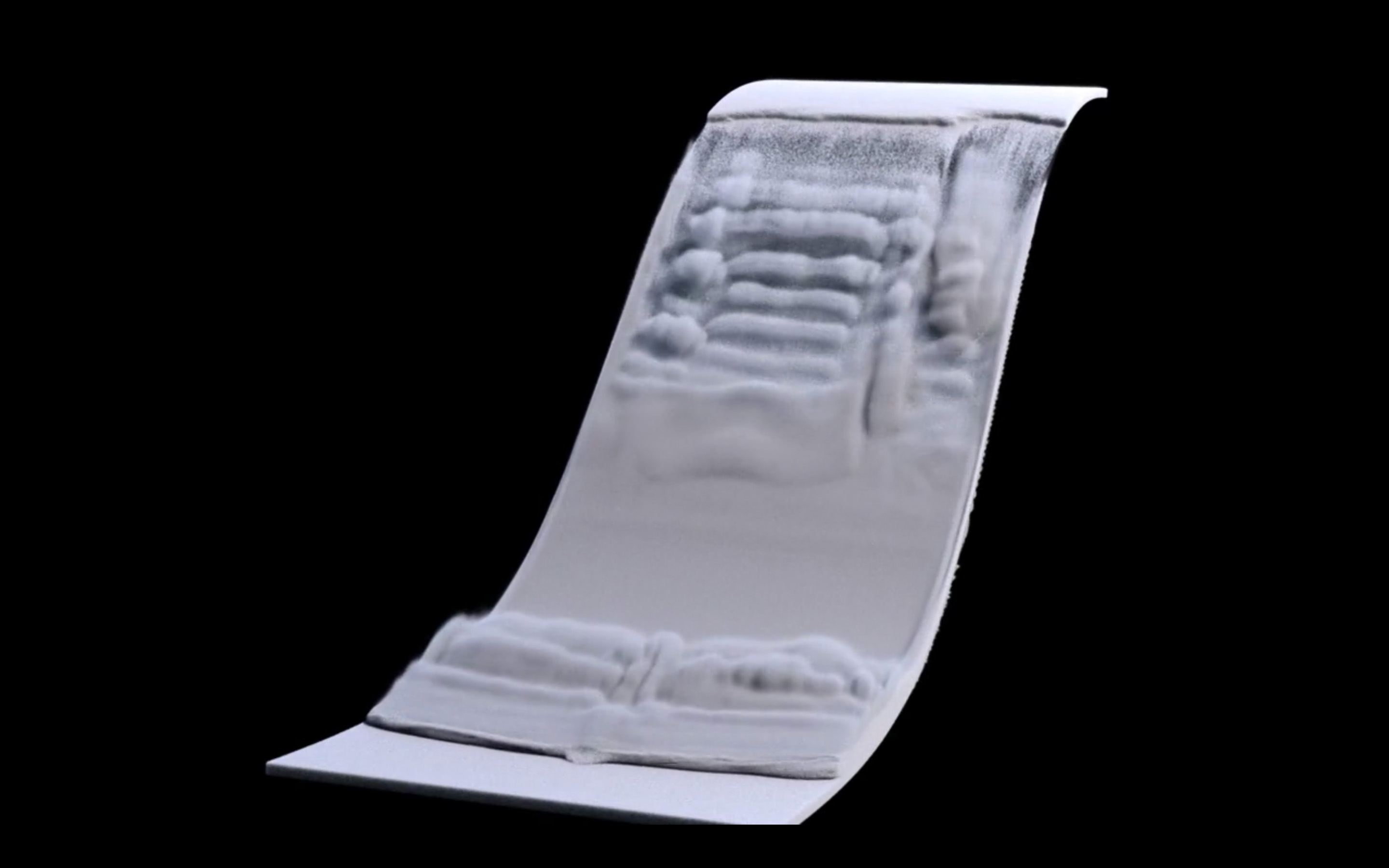}
\includegraphics[trim = 500 210 600 260, clip, width=0.33\linewidth]{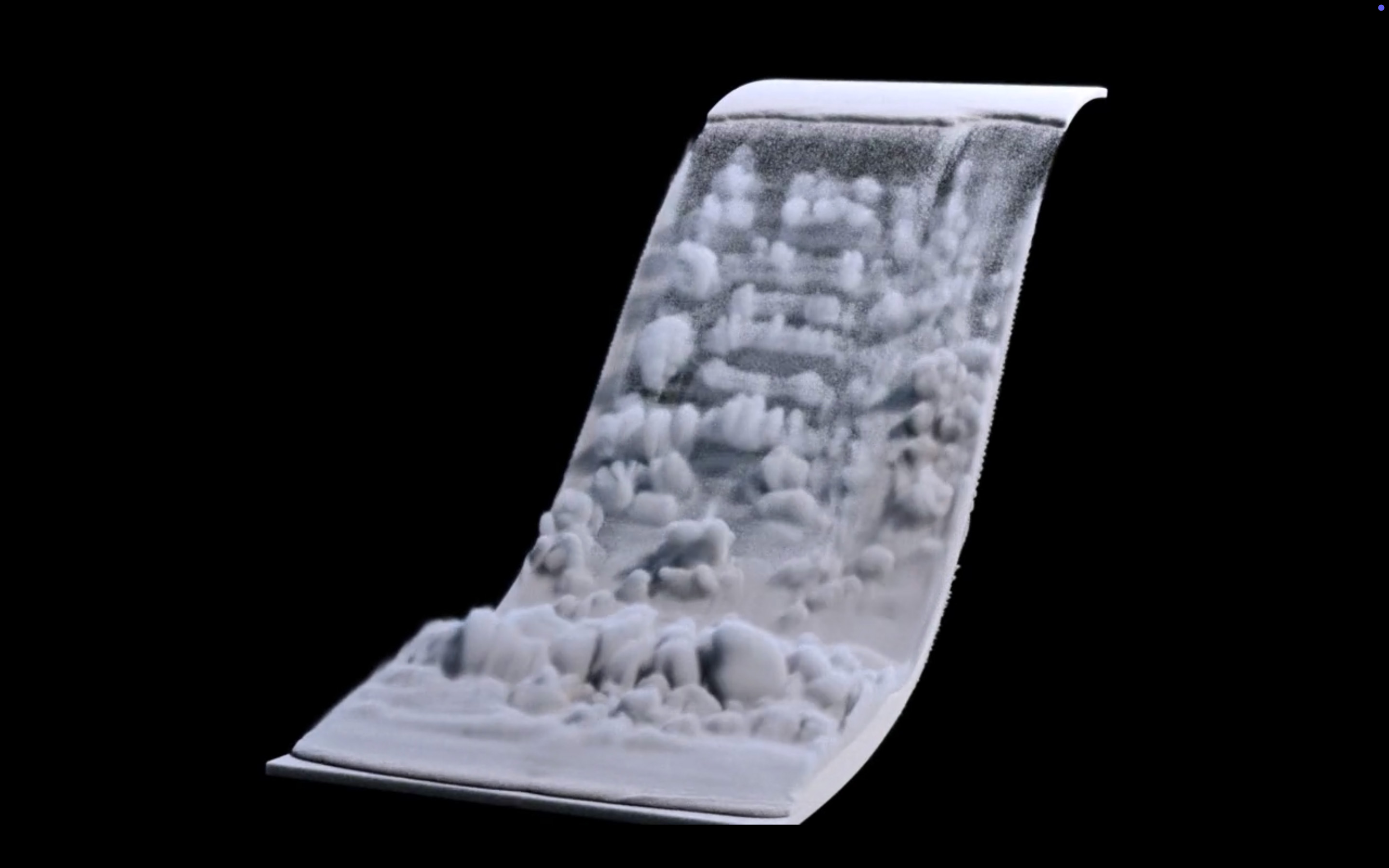}
\caption{\textbf{3D Avalanche on a slope.} A high-fidelity powder snow avalanche simulation descending a 3D curved slope.}
\label{fig:avalanche-3d}
\Description{}
\end{figure*}

\paragraph{Avalanche on the mountain} \autoref{fig:teaser} shows an avalanche in a real-world mountain scene, demonstrating the capability of our method in handling large-scale environments with complex terrain. With the snowpack initialized on the slopes, we simulate a snow avalanche triggered by a small perturbation. The adaptive refinement focuses computational resources on the snowpack and the flowing avalanche, capturing the intricate interactions between the snow particles and the underlying terrain. The simulation showcases the realistic behavior of the avalanche as it flows down the mountain, entraining powder snow and interacting with obstacles along its path. The large-scale environment and complex terrain highlight the efficiency and robustness of our adaptive LBM-MPM solver in simulating natural phenomena. \autoref{fig:avalanche-3d} demonstrates another 3D avalanche example. Compared to~\citet{Nascimento2025}, our method is capable of modeling dynamic crack propagation within snowpacks, interactions between buildings and dense snow, as well as powder-snow generation arising from snow entrainment processes.

\begin{figure*}[ht!]
\centering
\includegraphics[trim = 0 0 12cm 0, clip, width=0.33\linewidth]{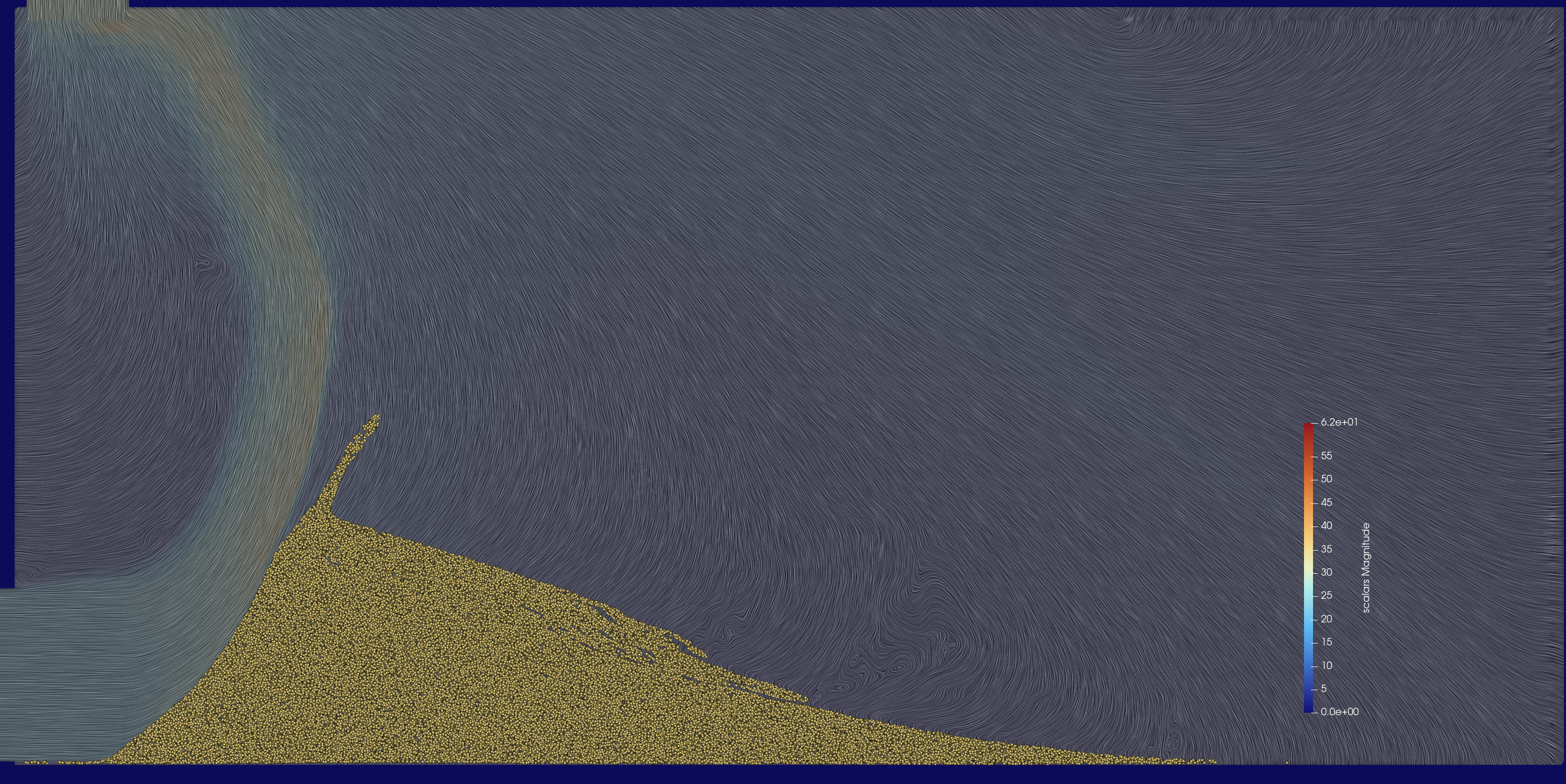}
\includegraphics[trim = 0 0 12cm 0, clip, width=0.33\linewidth]{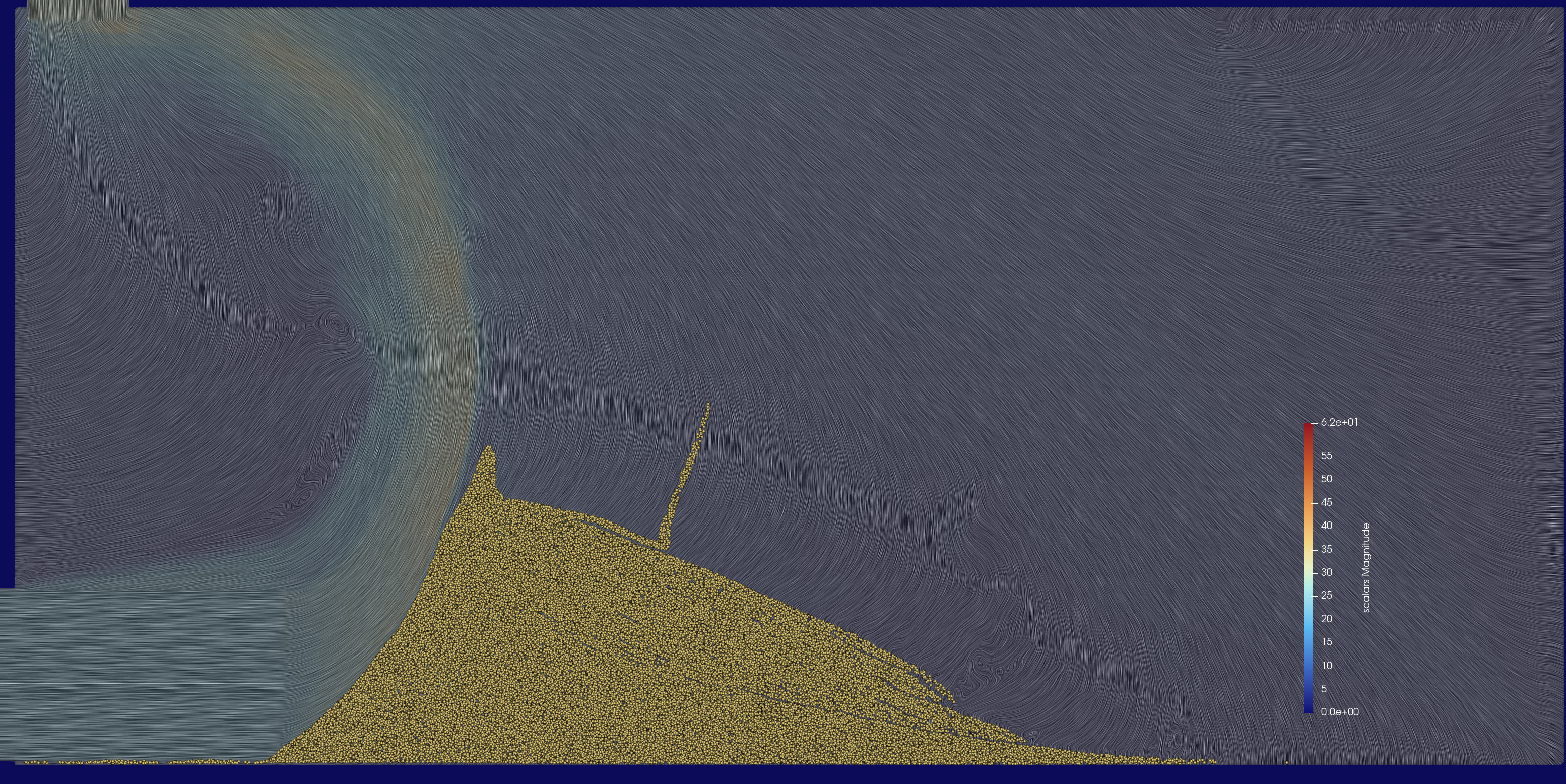}
\includegraphics[trim = 0 0 12cm 0, clip, width=0.33\linewidth]{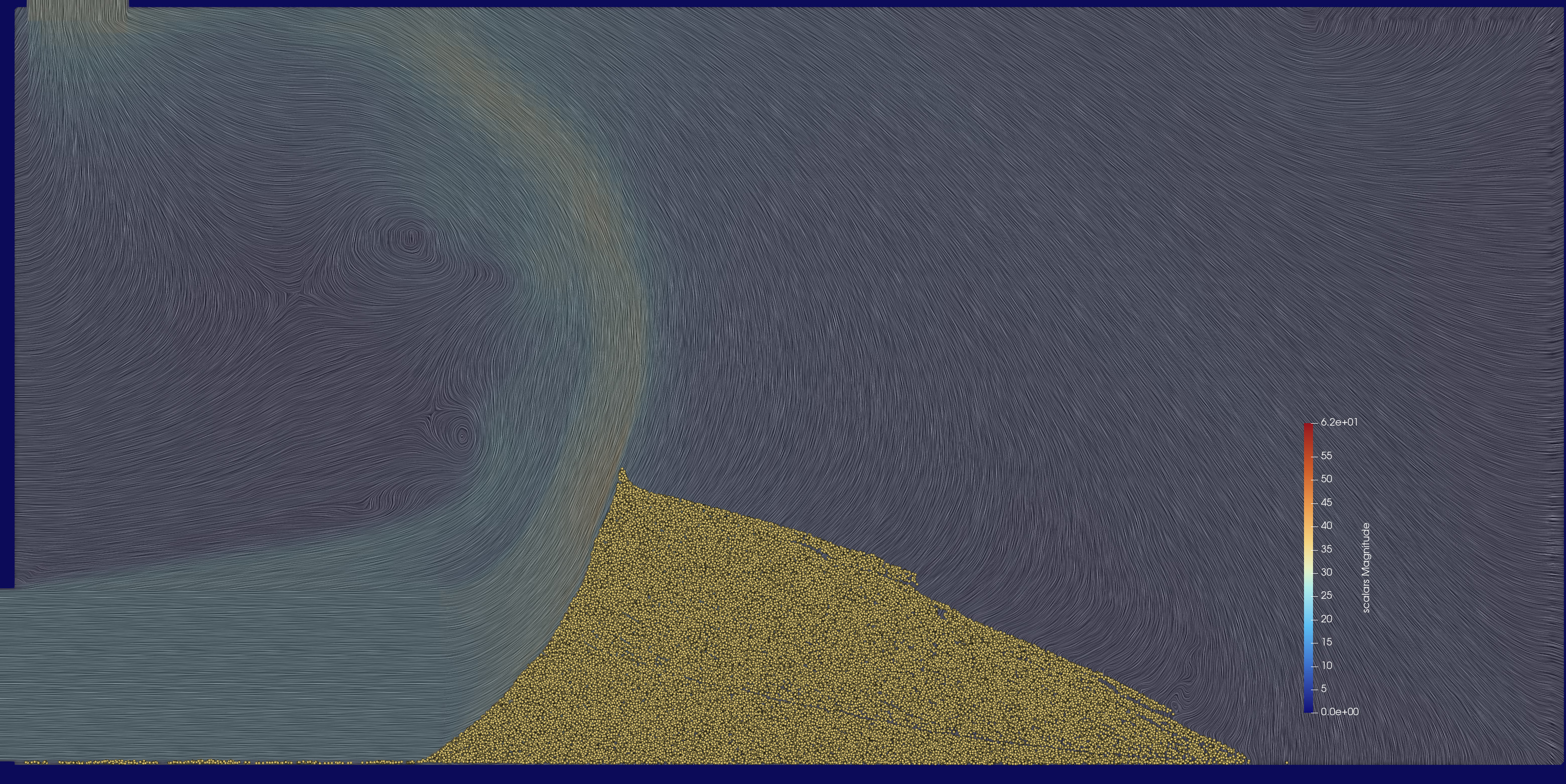}
\caption{\textbf{Dune migration}. A sand dune is posed in an open wind flow field, and migrates forward with the propelling of the wind. An open boundary is modeled with a convective boundary condition.}
\label{fig:dune}
\Description{}
\end{figure*}

\paragraph{Dune migration} 
\autoref{fig:dune} shows a pile of sand particles resting on the ground. As the wind blows from left to right, sand particles are lifted and transported by the airflow, leading to the gradual formation and migration of sand dunes. We adopt the convective boundary condition \cite{lou2013evaluation} at the wall boundaries, allowing the flow to naturally exit the boundaries. At an outlet boundary, we fetch the fluid speed $u_n$ of its nearest non-boundary neighbor in the direction of boundary normal, and solve the convective equation
$
\frac{\partial \mathcal{M}}{\partial t}+u_n\frac{\partial \mathcal{M}}{\partial n}=0
$
at the boundary cells. For the sand dune, we use the Drucker-Prager plastic model \cite{Klar2016Sand} with a volume correction \cite{Tampubolon2017MultiSpecies} to better capture sand motion. The adaptive refinement enables us to allocate computational resources to regions with high particle concentration and complex fluid-particle interactions, resulting in realistic dune shapes and movement patterns. 

\begin{figure*}[ht!]
\centering
\includegraphics[trim = 0 0 0 0, clip, width=\linewidth]{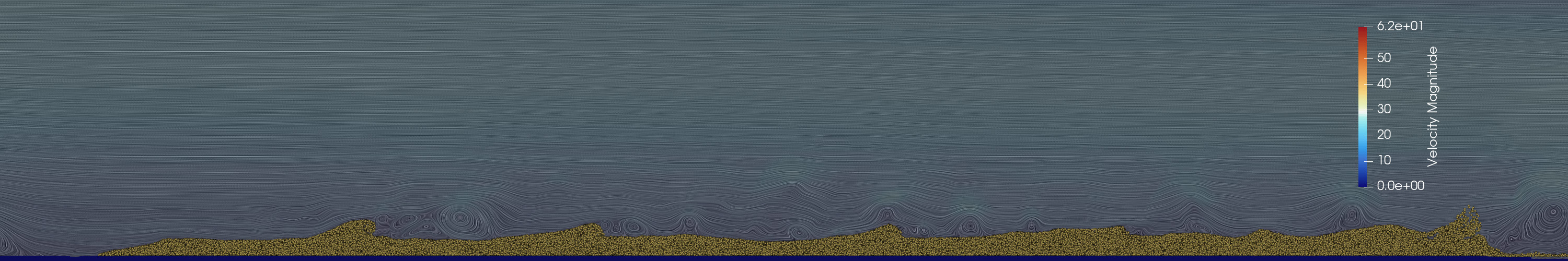}
\caption{\textbf{Sand Ripples.} This scenario illustrates the temporal evolution of a granular bed subjected to a shearing flow, showing the spontaneous emergence of sedimentary ripples.}
\label{fig:ripple-2d}
\Description{}
\end{figure*}

\paragraph{Sand ripples} 
Considering the friction of wind flow, we set wind velocity profile at the left boundary as a logarithmic distribution along the vertical direction~\cite{giudice2020},
$u_x(y)=u_0 \ln(1+\beta (y-y_0))$, where $u_0$ is the reference wind speed, $\beta$ is the wind shear coefficient, and $y_0$ is the characteristic height above the sand bed. The other boundaries are set to convective boundary conditions to approximate an open environment. \autoref{fig:ripple-2d} demonstrates a 2D example. As wind interacts with sand particles, small perturbations on the sand bed grow over time, leading to the emergence of ripple patterns. Compared with~\citet{Gao2018Mixture}, our framework additionally supports turbulent dusty flows interacting with MPM particles, enabled by our multi-level LBM solver, which allows the formation of sand ripple patterns.

\begin{figure*}[ht!]
\centering
\includegraphics[trim = 200 100 230 50, clip, width=0.498\linewidth]{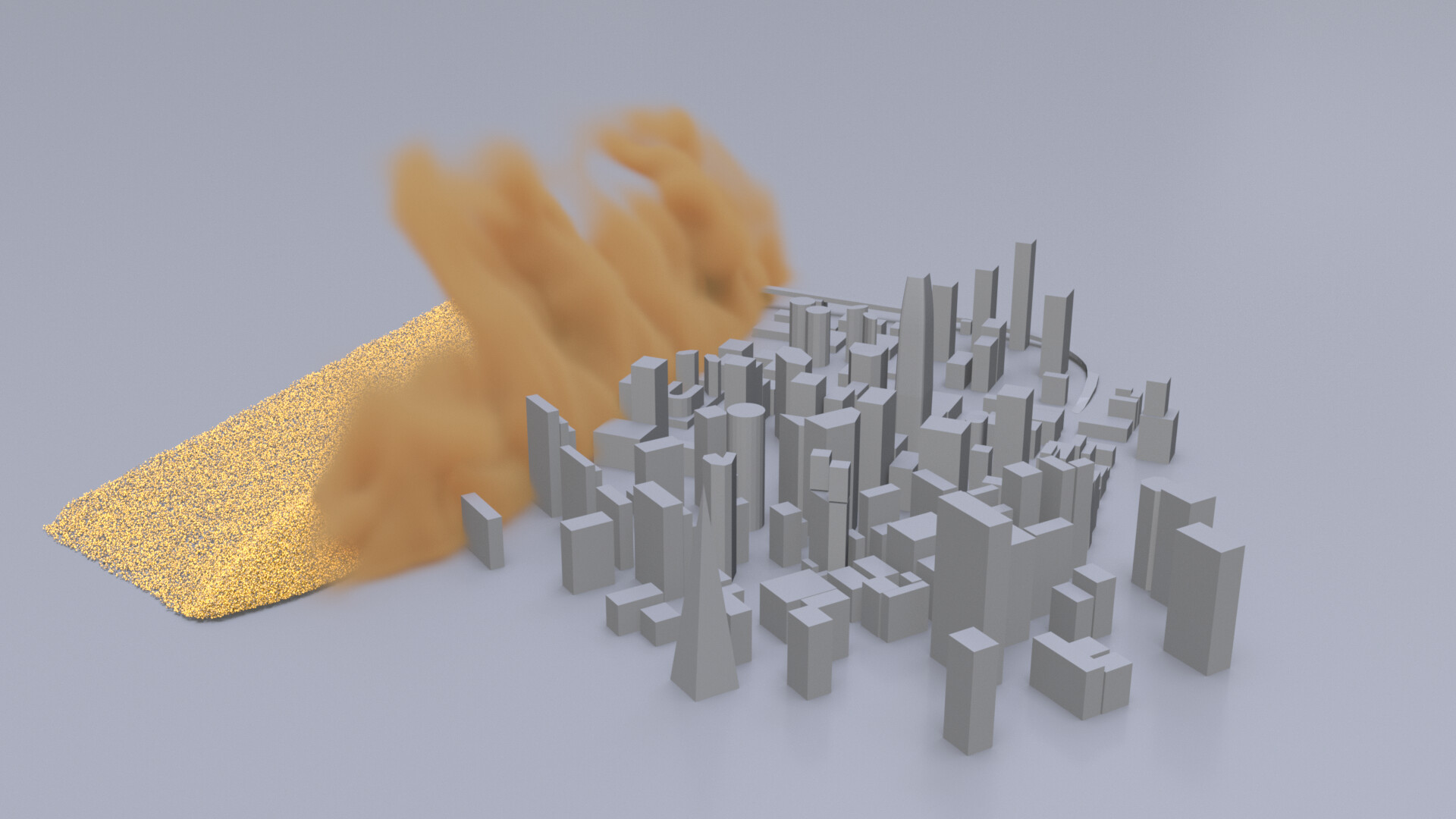}
\includegraphics[trim = 200 100 230 50, clip, width=0.498\linewidth]{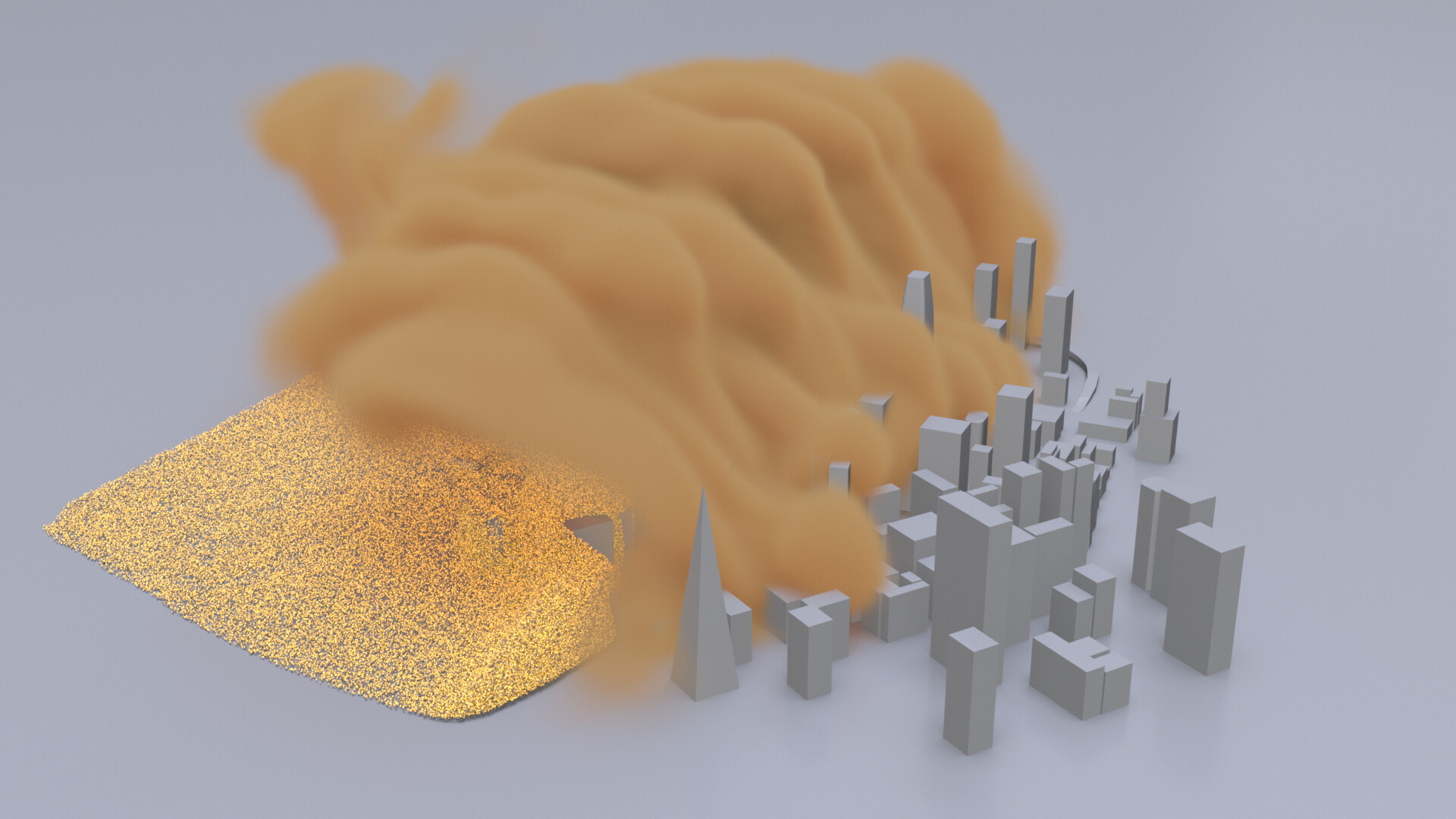}
\caption{\textbf{Sandstorm.} Large amount of dust is emitted during the dune migration driven by wind, forming a sand storm in front of the city.}
\label{fig:sandstorm}
\Description{}
\end{figure*}

\paragraph{Sandstorm} We further simulate the sandstorm in a procedural city scene, as shown in~\autoref{fig:sandstorm}. Initially, the sand dune is located near the city. The strong wind is pushing the sand dune forward, leading to a large sandstorm in front of the city. With our method, the turbulent sandstorm naturally emerges from dynamic sands, showcasing the potential for urban disaster prevention and mitigation.

\begin{figure*}[ht!]
\centering
\includegraphics[trim = 200 0 100 0, clip, height=0.36\linewidth]{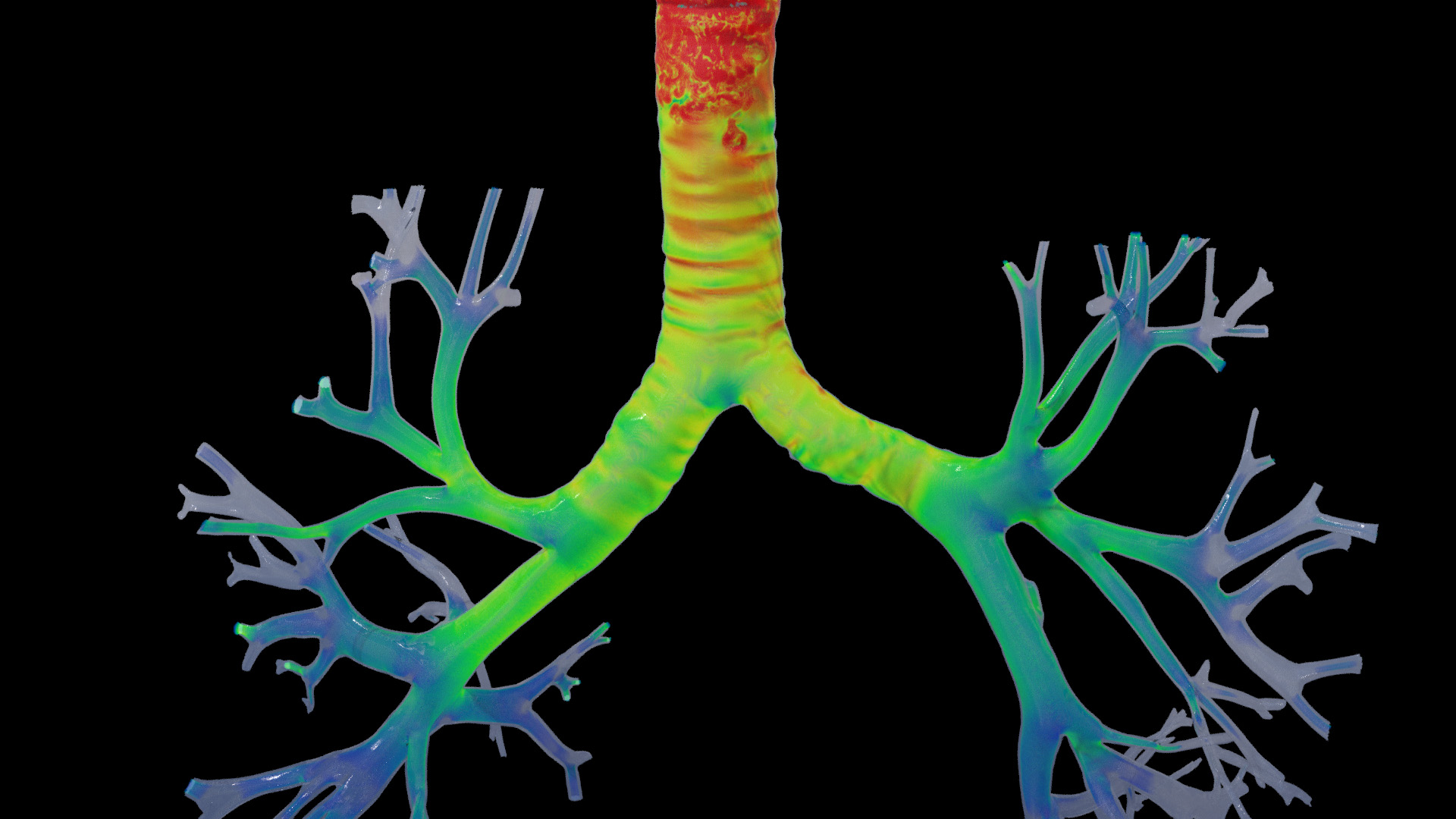}\hfill
\includegraphics[trim = 500 0 500 0, clip, height=0.36\linewidth]{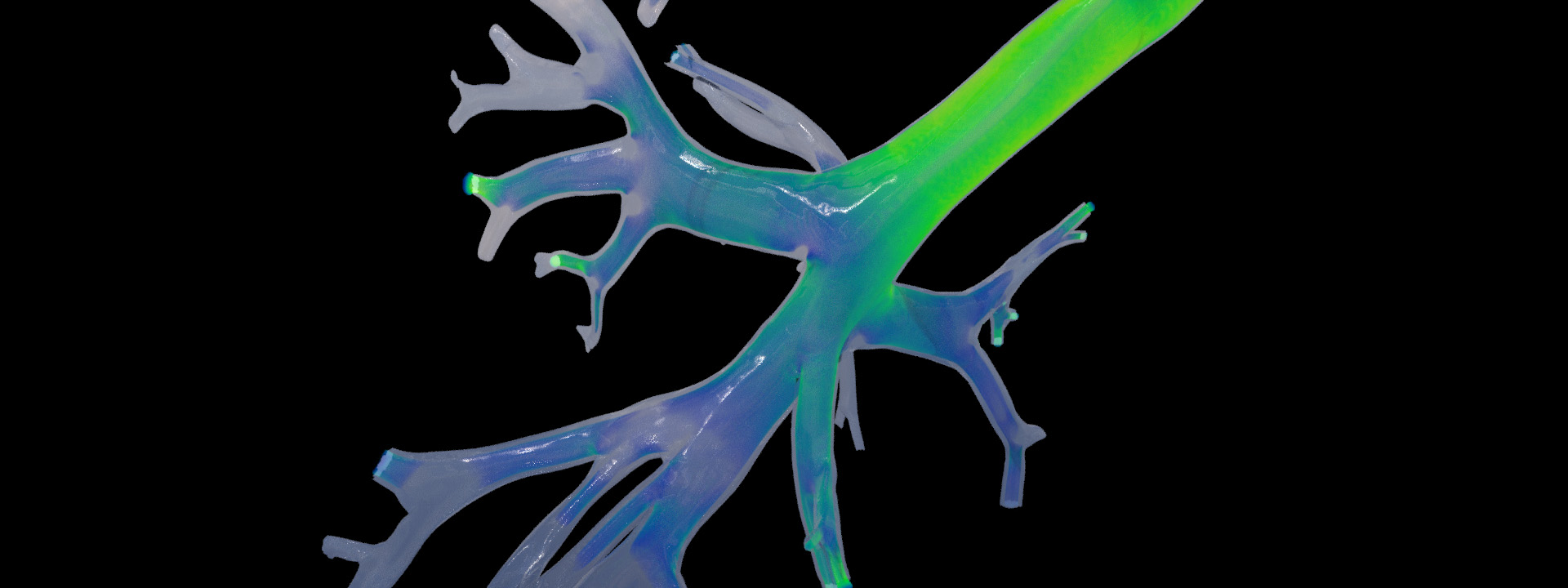}
\caption{\textbf{Flow in a bronchi.} A directional flow is injected into the model of bronchi with zoom-in on the right. Color represents the magnitude of vorticity. The injected flow can reach the end of the bronchi with an effective background resolution of $2048^3$. }
\label{fig:bronchi}
\Description{}
\end{figure*}

\paragraph{Flow in a bronchi}
\autoref{fig:bronchi} demonstrates a bronchial tree modeled as a series of branching tubes with varying diameters, representing the airways in the human respiratory system. We apply a pulsatile flow condition at the inlet to mimic the breathing cycle, while the outlet is open to the air. Since the diameter of the narrow branches is about $1/10$ that of the thick ones, we perform adaptive refinement based on the tube wall, achieving an extremely high effective resolution of $2048^3$ on a single GPU. The simulation captures the dynamic behavior of airflow through the bronchial tree, including vortex formation and mucus transport, showcasing the potential of our method in biomedical applications. 

\paragraph{Performance} \autoref{tab:stats} summarizes key statistics from our simulations. For the example shown in~\autoref{fig:bronchi}, uniform grids at such effective resolutions (up to $2048^3$) would far exceed the memory capacity of a single GPU. Our adaptive approach overcomes this limitation, achieving memory savings ranging from $5.8\times$ to $41.4\times$. This capability is critical in practice; for example, it enables the resolution of fine terminal branches in the Bronchi scene, which would otherwise be intractable on a single GPU. Moreover, the overhead introduced by our adaptive algorithm is negligible compared to the solver integration time (see supplementary document), ensuring high efficiency for large-scale coupled simulations. 
We further report performance results for 3D LBM–MPM coupling scenarios. Overall, thanks to the highly efficient adaptive LBM solver and the incremental updates of multi-level blocks, the primary performance bottleneck of the simulation remains the MPM stage, which accounts for approximately 69\%–89\% of the total runtime. This is expected, as MPM involves computationally intensive particle state updates and information exchange between particles and grid.

\begin{table}[ht]
\vspace{-0.25em}
\caption{Statistics of 3D examples. Values in parentheses represent the ratio of effective resolution to the actual number of cells in tiles.}\label{tab:stats}
\vspace{-1em}
\scalebox{0.9}
	{ 
\centering
\begin{tabular}{l c c c c c}
\hline
Figure & Effective Res & \#Particles & Max \#Tiles & Time/Step \\
\hline
\autoref{fig:teaser} & $1536\times 768\times 384$ & 55.8M & 733k (9.4$\times$)  & 5.25s \\
\autoref{fig:avalanche-3d} & $800\times 400\times 400$ & 63.1M & 321k (6.2$\times$)  & 1.65s \\
\autoref{fig:sandstorm} & $512\times 128\times 512$ & 1.01M & 90.1k (5.8$\times$)  & 0.12s \\
\autoref{fig:bronchi} & $2048\times 2048\times 2048$ & -- & 3.24M (41.4$\times$)  & 1.46s \\
\hline
\end{tabular}
}
\vspace{-0.75em}
\end{table}
\section{conclusion}

We have presented a unified framework for large-scale fluid–granular simulations that couples a multi-level, moment-based LBM solver with MPM. Our approach achieves both high physical fidelity and computational efficiency by enabling adaptive multi-resolution simulations, consistent moment rescaling across different resolutions, and dynamic GPU-based grid maintenance. Through accurate two-way coupling between fluids and granular media, our framework can simulate a wide range of phenomena, from snow avalanches to sandstorms, at scales and resolutions previously impractical. This work enables more realistic and efficient simulations of multi-material flows in graphics, engineering, and virtual environments.

\paragraph{Limitations}
Our method has several limitations that suggest promising directions for future work. First, our multi-level alignment strategy relies on overlapping 5–8 cells between adjacent levels to ensure stability and accuracy; while effective, this overlap introduces nontrivial computational overhead and could be improved with more efficient alignment or constraint formulations. Second, although our solver supports multiphase flows, the coupling between air, water, and granular materials remains limited, restricting its ability to model fully coupled air–water–granular interactions.

\begin{acks}

\end{acks}

\clearpage
\newpage

\bibliographystyle{ACM-Reference-Format}
\bibliography{bibliography}

\clearpage
\newpage

\end{document}